\documentclass[twocolumn,showpacs,preprintnumbers,amsmath,amssymb]{revtex4}
\usepackage{epsfig}
\usepackage{graphicx}
\usepackage{dcolumn}
\usepackage{bm}
\usepackage{amsthm}
\usepackage{amsfonts}

\newcommand{\ket}[1]{\mbox{$ | #1 \rangle $}}
\newcommand{\bra}[1]{\mbox{$ \langle #1 | $}}

\begin{document}

\title{Effect of detector dead-times on the security evaluation of differential-phase-shift
quantum key distribution against sequential attacks}

\author{Marcos Curty$^{1}$, Kiyoshi Tamaki$^{2,3}$ and Tobias
Moroder$^{4,5}$} \affiliation{ $^1$
ETSI Telecomunicaci\'on, 
University of Vigo, Campus Universitario, 36310 Vigo, Spain \\
$^2$ NTT Basic Research Laboratories, NTT Corporation, 3-1
Morinosato Wakamiya Atsugi-Shi, Kanagawa, 243-0198, Japan \\
$^3$ CREST, JST Agency, 4-1-8 Honcho, Kawaguchi, Saitana,
332-0012,
Japan \\
$^4$ Institute for Quantum Computing, University of Waterloo, Waterloo, ON, N2L 3G1, Canada \\
$^5$ Quantum Information Theory Group, Institut f\"ur Theoretische
Physik I, and Max-Planck Research Group, Institute of Optics,
Information and Photonics, University of Erlangen-N\"urnberg,
91058 Erlangen, Germany}

\date{\today}

\begin{abstract}
We investigate limitations imposed by detector dead-times on the
performance of sequential attacks against a
differential-phase-shift (DPS) quantum key distribution (QKD)
protocol with weak coherent pulses. In particular, we analyze
sequential attacks based on unambiguous state
discrimination of the signal states emitted by the source and we
obtain ultimate upper bounds on the maximal distance achievable
by a DPS QKD scheme both in the so-called trusted and untrusted
device scenarios, respectively.  
\end{abstract}


\maketitle

\section{INTRODUCTION}

Quantum key distribution (QKD) \cite{gisin_rev_mod} is a
technique that allows two parties (usually called Alice and Bob)
to generate a secret key despite the computational and
technological power of an eavesdropper (Eve) who interferes with
the signals. Together with the Vernam cipher \cite{vernam}, QKD
can be used for unconditionally secure data transmission.

The first complete QKD scheme was introduced by Bennett and
Brassard in 1984 (BB84 for short) \cite{BB84}. An unconditional 
security proof for the whole protocol has been given in
Ref.~\cite{Mayers98}. After the first demonstration of the
feasibility of this scheme \cite{Bennett92}, several
long-distance implementations of QKD have been realized in the
last years (see, for instance, Ref.~\cite{Marand95} and references
therein). However, these practical approaches differ in many
important aspects from the original theoretical proposal, since
it demands technologies that are beyond our present
experimental capability. Especially, the signals emitted by the
source, instead of being single-photons, are usually weak
coherent pulses (WCP) with typical average photon numbers of
$0.1$ or higher. 
This fact, together with the considerable
attenuation introduced by quantum the channel and the noise introduced 
by the detectors, jeopardize the security of the protocol and lead to
limitations of rate and distance that can be covered by these
techniques \cite{Huttner95,Norbert00}. A positive security proof
against all individual attacks, even with practical signals, has
first been given in Ref.~\cite{Norbert_individual}, while a
complete proof of the unconditional security of this scheme in a
realistic setting has been provided in Refs.~\cite{inamori,inamori2}. 
This
means that, despite practical restrictions, with the support of
the classical information techniques (error correction and privacy amplification) 
used in the key distillation
phase, it is still possible to obtain a secure secret key.

The main security threat of QKD protocols based on WCP arises from the fact
that some signals contain more than one photon prepared in the same
polarization state. Now, Eve can perform, for instance, the
so-called {\it Photon Number Splitting} (PNS) attack on the
multi-photon pulses \cite{Huttner95}. This attack provides Eve
with full information about the part of the key generated from
the multi-photon signals, without causing any disturbance in the
signal polarization. As a result, it turns out that the BB84
protocol with WCP can give a key generation rate of order
$O(\eta^2)$, 
where $\eta$ denotes the transmission 
efficiency of the quantum channel \cite{inamori,inamori2}.

To obtain higher secure key rates over longer distances,
different QKD schemes, that are robust against the PNS attack, have been
proposed in recent years. One of these schemes is the so-called
decoy-states \cite{decoy_t,decoy_e}, where Alice varies at random
the mean photon number of the signal states sent to Bob by using
different intensity settings. This technique delivers a key
generation rate of order $O(\eta)$ \cite{decoy_t,decoy_e}. Another
possibility is based on the transmission of two non-orthogonal
coherent states together with a strong reference pulse
\cite{ben92}. This scheme has been analyzed in detail in
Ref.~\cite{koashi04}, where it was confirmed that also in this
scenario the secure key rate is of order $O(\eta)$. Finally,
another possible approach is to use a differential-phase-shift
(DPS) QKD protocol
\cite{dpsqkd,dpsqkd2,dpsqkd_exp1,dpsqkd_exp2,dpsqkd_exp2b,dpsqkd_exp3}. In
this scheme Alice sends to Bob a train of WCP whose
phases are randomly modulated by $0$ or $\pi$. On the receiving
side, Bob measures out each incoming signal by means of an
interferometer whose path-length difference is set equal to the
time difference between two consecutive pulses. In this last 
case, however, a
secure key rate of order $O(\eta)$ has only been proven so far
against a special type of individual attacks where Eve acts and measures 
{\it photons} individually, rather than {\it signals}
\cite{dpsqkd2}, and also against a particular class of collective attacks where 
Eve attaches ancillary systems to each pulse or to each 
pair of successive pulses \cite{cyril_new}. While a complete security proof of DPS QKD 
against the most general attack is still missing, recently it has been 
shown that sequential attacks already impose strong 
restrictions on the performance of this QKD scheme with
WCP \cite{dpsqkd2,curty_dps,tsurumaru_dps}. For instance, it was 
proven in
Refs.~\cite{curty_dps,tsurumaru_dps} that the DPS QKD experiments
reported in Refs.~\cite{dpsqkd_exp1,dpsqkd_exp2} are insecure
against this type of attacks. Basically,
a sequential attack consists of Eve measuring out every
signal state emitted by Alice and, afterwards, she prepares new signal states,
depending on the results obtained, that are given to Bob.
Whenever Eve obtains a predetermined number of consecutive
successful measurement outcomes, then she prepares
a train of non-vacuum signal states that is forwarded to Bob. Otherwise,
Eve sends, for instance, vacuum signals to Bob to avoid errors. Sequential attacks constitute 
a special type of intercept-resend attacks \cite{jahma01,Felix01,curty05} 
and, therefore, they provide ultimate upper bounds on the performance
of QKD schemes \cite{Curty04}. 

In discussions within
the scientific community one often hears, however, that the 
security analysis presented in Refs.~\cite{curty_dps,tsurumaru_dps} might 
have overestimated the strength that sequential attacks have against a DPS QKD 
protocol. This conjecture 
is justified because the 
sequential attacks studied so far in the literature have not considered the effect of 
Bob's detectors dead-time. As a 
result, the probability that each non-vacuum signal state, within a 
train of them, sent by Eve 
contributes to the sifted key does not depend on whether the previous 
signal states in the train already produced a click on Bob's detection 
apparatus or not. This suggests that such analysis might overestimate the 
number of Bob's detected events that originates from the non-vacuum
signal states sent by Eve and, therefore, it might deliver shorter 
secure distances.

The aim of this paper is to investigate limitations imposed by 
Bob's detectors dead-time on the
performance of sequential attacks against a DPS QKD protocol 
\cite{note1}. For that, we shall analyze 
sequential attacks based on unambiguous state discrimination 
(USD) of the signal states emitted by Alice
\cite{curty_dps,tsurumaru_dps,usd,chef,jahma01}. When Eve
identifies unambiguously a signal state, then she considers this
result as successful. Otherwise, she considers it a failure.
We shall consider two possible scenarios for our analysis. The
first one, so-called {\it untrusted device scenario}, arises from
a conservative definition of security, {\it i.e.}, we shall assume that
Eve can control some imperfections in Alice and Bob's devices ({\it
e.g.}, the detection efficiency, the dark count probability, 
and the dead-time of Bob's detectors), together with the losses
in the quantum channel, and she exploits them to obtain maximal
information about the shared key. In the second scenario,
so-called {\it trusted device scenario}, we shall consider that Eve
cannot modify the actual detection devices employed by Alice and
Bob. That is, the legitimate users have complete knowledge
about their detectors, which are fixed by the actual experiment.
The main motivation to
study this scenario is that, from a practical point of view, it
constitutes a reasonable description of a realistic situation,
where Alice and Bob can limit Eve's influence on their apparatus
by some counterattack techniques \cite{Note6}.

A different QKD scheme, but also related to a DPS QKD protocol, has been 
proposed recently in Ref.~\cite{stucki_new}. (See also Ref.~\cite{gisin_new}.)
However, since the abstract signal structure of this protocol
is different from the one of a DPS QKD scheme, the analysis 
contained in this paper does not apply to that scenario. Sequential 
attacks against the QKD protocol introduced in Ref.~\cite{stucki_new}
have been investigated in Ref.~\cite{bran_new}, while its security 
against a particular class of collective attacks has been studied 
in Ref.~\cite{cyril_new}.   

The paper is organized as follows. In Sec.~\ref{sec_1} we
describe in more detail a DPS QKD protocol. Then, in
Sec.~\ref{seqattack}, we present sequential attacks against this
QKD scheme. Section~\ref{susda} includes the analysis for the
untrusted device scenario. Here we obtain an upper bound on the
maximal distance achievable by a DPS QKD protocol as a function of
the error rate in the sifted key, the mean photon-number of
Alice's signal states and the dead-time of Bob's detectors.
Similar results are derived in Sec.~\ref{medattack}, now for the
trusted device scenario. Finally, Sec.~\ref{CONC} concludes the
paper with a summary. The manuscript includes as well several 
appendices with additional calculations.

\section{DIFFERENTIAL-PHASE-SHIFT (DPS) QKD}\label{sec_1}

\begin{figure}
\begin{center}
\includegraphics[scale=1.1]{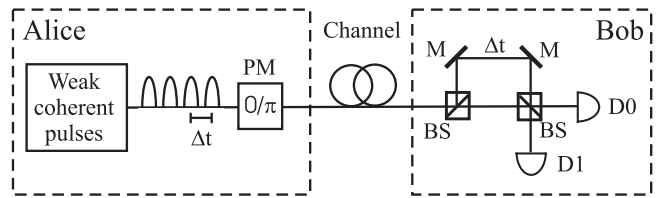}
\end{center}
\caption{Basic setup of a DPS QKD scheme. PM denotes a phase
modulator, BS, a $50:50$ beam splitter, M, a mirror, D0 and D1 are
two photon detectors and $\Delta{}t$ represents the time
difference between two consecutive pulses. \label{dpsqkd}}
\end{figure}

The setup is illustrated in Fig.~\ref{dpsqkd}
\cite{dpsqkd,dpsqkd2,dpsqkd_exp1,dpsqkd_exp2,dpsqkd_exp2b,dpsqkd_exp3}.
Alice prepares first a train of coherent states $\ket{\alpha}$
and, afterwards, she modulates, at random and independently every
time, the phase of each pulse to be $0$ or $\pi$. As a result,
she produces a random train of signal states $\ket{\alpha}$ or
$\ket{-\alpha}$ that are sent to Bob through the quantum channel.
On the receiving side, Bob uses a $50:50$ beam splitter to divide
the incoming pulses into two possible paths and then he recombines
then again using another $50:50$ beam splitter. The time delay
introduced by Bob's interferometer is set equal to the time
difference $\Delta{}t$ between two pulses. Whenever the relative
phase between two consecutive signals is $0$ ($\pm\pi$) only the
photon detector $D0$ ($D1$) may produce a ``click" (at least one
photon is detected). For each detected event, Bob records the
time slot where he obtained a click and the actual detector that
fired.

Once the quantum communication phase is completed, Bob uses a
classical authenticated channel to announce the time slots
where he obtained a click, but he does not reveal which detector
fired each time. From this information provided by Bob,
together with the knowledge of the phase value used to modulate
each pulse, Alice might infer which photon detector had clicked at Bob's
side each given time. Then, Alice and Bob can agree, for instance,
to select a bit value ``0" whenever the photon detector $D0$
fired, and a bit value ``1" if the detector $D1$ clicked. In an
ideal scenario, Alice and Bob end up with an identical string of
bits representing the sifted key. Due to the noise
introduced by the quantum channel, together with possible
imperfections of Alice and Bob's devices, however, the sifted
key typically contains some errors. Then, Alice and Bob perform
error-correction to reconcile the data and privacy amplification
to decouple the data from Eve. (See, for instance,
Ref.~\cite{gisin_rev_mod}.)


\section{Sequential attacks against DPS QKD}
\label{seqattack}

A sequential attack can be seen as a special type of
intercept-resend attack \cite{dpsqkd2,curty_dps,tsurumaru_dps}.
First, Eve measures out every coherent state emitted by Alice with
a detection apparatus located very close to the sender.
Afterwards, she transmits each measurement result through a
lossless classical channel to a source close to Bob. Whenever Eve
considers a sequence of measurement outcomes {\it successful},
this source prepares a new train of signal states that is forwarded to Bob.
Otherwise, Eve typically sends vacuum signals to Bob to avoid errors.
Whether a sequence of measurement results is considered to be
successful or not, and which type of non-vacuum signal states Eve sends
to Bob, depends on Eve's particular eavesdropping strategy and on
her measurement device. Sequential attacks transform the original
quantum channel between Alice and Bob into an entanglement
breaking channel \cite{Horodecki03} and, therefore, they do not
allow the distribution of quantum correlations needed to
establish a secret key \cite{Curty04}.

Let us begin by introducing Eve's measurement apparatus. As mentioned
previously, we shall consider that Eve realizes USD \cite{usd,chef} 
of each signal
state sent by Alice. That is, whenever she obtains a conclusive result 
then it is guaranteed that the result is always correct. 
In order to do that, we will assume 
that Eve has always access to a local oscillator that is
phase-locked to the coherent light source employed by Alice
\cite{Note2}.
Whenever Eve identifies unambiguously a
predetermined number of consecutive signal states, {\it i.e.},
she determines without error whether each signal state is
$\ket{\alpha}$ or $\ket{-\alpha}$, she considers this sequence of
measurement outcomes successful. Otherwise she considers it a
failure \cite{Note3}. We define the integer parameter $M_{min}$ as
the minimum number of consecutive USD successful results that Eve
needs to obtain in order to consider the sequence of measurement
outcomes successful. More precisely, if $k\geq{}0$ denotes the
total number of consecutive USD successful outcomes obtained by
Eve before she obtains an inconclusive result, then, whenever
$k>M_{min}$, Eve prepares a new train of signal states, that we shall denote
as $\rho_e^k$, together with some vacuum states for the
inconclusive result, and she sends these signals to Bob. 
The precise definition of the quantum state $\rho_e^k$ will be 
introduced later on, 
since it will depend
on whether we consider the untrusted or the trusted device
scenario, respectively. The reason to append {\it some} vacuum states 
to each train of signal states $\rho_e^k$ is also closely related 
to the eavesdropping strategy of these two possible cases.
The main idea behind this procedure is to guarantee that whenever Bob 
obtains a click on his detection apparatus then he cannot obtain 
any other click afterwards during 
a period of time at least equal to the dead-time of his detectors.
That is, these vacuum states sent by Eve will allow her to reproduce the dead-time
of Bob's detectors, whose influence on the security evaluation of a DPS QKD 
protocol is the main focus of this paper.  
For simplicity, let us assume for the moment that Eve sends to 
Bob $1+d$ vacuum 
states together with each train of signal states $\rho_e^k$ in order 
to achieve this goal, while the precise value of the 
parameter $d$ will be given for the untrusted (trusted) device scenario
in Sec.~\ref{susda} (Sec.~\ref{medattack}). 
On the other hand, if $k<M_{min}$ Eve
sends to Bob $k+1$ vacuum states, 
where the last vacuum state corresponds to Eve's inconclusive 
result. The case $k=M_{min}$ deserves
special attention. We shall consider that in this situation Eve
employs a probabilistic strategy that combines the two previous
ones. In particular, we assume that Eve sends to Bob the signal
state $\rho_e^{M_{min}}$, together with $1+d$ vacuum states, with
probability $q$ and, with probability $1-q$, she sends to Bob
$M_{min}+1$ vacuum states. That is, the parameter $q$ allows Eve
to smoothly fit her eavesdropping strategy to the observed data
\cite{curty_dps}. Moreover, in order to simplify our calculations, 
we define the
integer parameter $M_{max}>M_{min}$ as the maximum number of
consecutive USD successful results that Eve can obtain in order
to send to Bob a train of signal states. That is, whenever Eve obtains
$M_{max}$ consecutive USD successful outcomes then she discards
the next measurement outcome and directly sends to Bob the quantum
state $\rho_e^{M_{max}}$ together with $1+d$ vacuum states for the
discarded measurement result.

Let $p$ denote the probability that Eve obtains an USD successful
result per signal state sent by Alice. It has the following 
form \cite{usd}
\begin{equation}\label{spusd}
p=1-\vert\bra{\alpha}-\alpha\rangle\vert=1-\exp{(-2\mu_\alpha)},
\end{equation}
where $\mu_\alpha$ represents the mean photon-number of Alice's
signal states, {\it i.e.}, $\mu_\alpha=\vert\alpha\vert^2$.

\begin{figure}
\begin{center}
\includegraphics[scale=1.1]{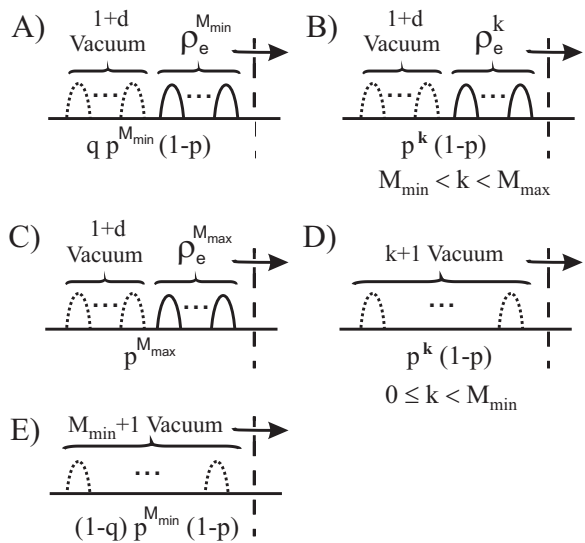}
\end{center}
\caption{Possible signal states that Eve sends to Bob 
together with their a priori probabilities.
The arrow indicates the transmission direction.
\label{strategy1_meas}}
\end{figure}

We shall denote with $p_s(k)$ the probability that Eve sends to
Bob a train of signal states $\rho_e^k$, together with $1+d$ vacuum states.
This probability can be written as
\begin{equation}\label{ps_k}
p_s(k) = \left\{ \begin{array}{ll} q p^{M_{min}}(1-p) &
\textrm{if $k=M_{min}$}\\
p^k(1-p) & \textrm{if $M_{min}<k<M_{max}$}\\
p^{M_{max}} & \textrm{if $k=M_{max}$}\\
0 & \textrm{otherwise,}
\end{array} \right.
\end{equation}
with $p$ given by Eq.~(\ref{spusd}).
Similarly, we shall denote with $p_v(k)$ the probability that Eve sends to Bob
$k+1$ vacuum states. This probability is given by
\begin{equation}\label{pv_k}
p_v(k) = \left\{ \begin{array}{ll}
p^k(1-p) & \textrm{if $0\leq{}k<M_{min}$}\\
(1-q)p^{M_{min}}(1-p) & \textrm{if $k=M_{min}$}\\
0 & \textrm{otherwise.}
\end{array} \right.
\end{equation}
We illustrate all these possible cases in Fig.~\ref{strategy1_meas}, where
we also include the different a priori probabilities to be in each of these
scenarios.

Next we analyze in detail the influence that Bob's
detectors dead-time has on the performance of the sequential
attack introduced above. The goal is
to find an expression for the gain, {\it i.e.}, the probability
that Bob obtains a click per signal state sent by Alice, together
with the quantum bit error rate (QBER) introduced by Eve, in both 
the untrusted and the trusted device scenarios, respectively.  

\section{Untrusted device scenario}\label{susda}

Here we shall consider that Eve can always control some imperfections in Alice and
Bob's devices together with the 
quantum channel. Especially, we shall assume that Eve can always replace Bob's
imperfect detection apparatus by an ideal one in order to exploit
its 
detection efficiency, together with the dark count
probability and the dead-time of his detectors, 
to obtain maximal
information about the shared key. Of course, to guarantee that Eve's presence
remains unnoticeable to the legitimate users, Eve needs to send Bob
signal states that can reproduce the statistics 
that Alice and Bob expect after their measurements. For this, we shall
consider the standard version of a DPS QKD
protocol, where Alice and Bob only monitor the raw bit rate (before the key
distillation phase) together with the time instances in which Bob
obtains a click.

The main limitation on the type of signal states that Eve can send to Bob in this 
scenario arises from the dead-time of Bob's detectors. In order to simplify our analysis 
we shall assume that both detectors $D0$ and $D1$ in 
Fig.~\ref{dpsqkd} are indistinguishable, 
{\it i.e.}, their dark count rate, quantum efficiency and dead-time, are equal.
Moreover, we 
shall consider a conservative scenario where every time that one of these 
detectors clicks, then both detectors do not respond 
to any other incident photon during a period of time 
equal to the dead-time, {\it i.e.}, we shall assume that after a click 
both detectors suffer {\it simultaneously}
from a dead time. This is a key assumption underlying the whole analysis presented 
in this paper. In the 
experimental setup employed in 
Refs.~\cite{dpsqkd_exp1,dpsqkd_exp2,dpsqkd_exp2b,dpsqkd_exp3} both detectors $D0$ and 
$D1$ are connected to the same Time Interval Analyser (TIA) that also has 
a dead-time which is typically much higher than the dead-time of the detectors. 
Whenever one of these detectors clicks then the TIA does not respond to any other click event 
during a period of time equal to its dead-time. In this situation, however, 
if $D0$ and $D1$ do not suffer simultaneously from a dead-time, then
the effect of the TIA can be understood as just blocking some of the output 
signals coming from the two detectors. As a result, the raw key obtained by Alice and 
Bob could contain correlations between different bits which might be known to 
some extend to Eve. For instance, if one detector clicks, and the clock 
frequency of the system is high enough, then, because of its dead-time, 
it is more probable that the next click comes from the other detector. 
This last scenario is beyond the scope of this paper and 
the analysis will be presented somewhere else.   
 
\begin{figure}
\begin{center}
\includegraphics[scale=1.1]{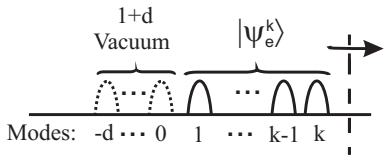}
\end{center}
\caption{Labeling convention for the $k$ temporal modes of the
signal state $\ket{\psi_k}$ given by Eq.~(\ref{signal_uds}) followed by $1+d$
vacuum states. The
arrow indicates the transmission direction. \label{ordering}}
\end{figure}

In the sequential attack introduced in Sec.~\ref{seqattack}, 
Eve sends to Bob only two possible
classes of signal states: a state $\rho_e^k$ followed by $1+d$
vacuum states, or a train of $k+1$ vacuum signals. This means, in
particular, that Bob can only obtain clicks in his detection apparatus 
when he receives a signal state
$\rho_e^k$. To be able to mimic the dead-time of Bob's detectors, therefore, Eve needs to 
select each state $\rho_e^k$ such that it can 
produce only one click on Bob's side within a dead-time period. For that, Eve chooses $\rho_e^k$
containing only one photon distributed among $k$ temporal modes. These 
modes correspond to $k$ consecutive pulses sent by Alice, {\it i.e.}, 
the time difference between two consecutive temporal modes in
$\rho_e^k$ is set equal to the time difference $\Delta{}t$
between two consecutive pulses sent by Alice. More precisely, we shall consider 
that $\rho_e^{k}$ denotes a pure state $\ket{\psi_e^k}$ ({\it i.e.}, 
$\rho_e^{k}=\ket{\psi_e^k}\bra{\psi_e^k}$) given by
\begin{equation}\label{signal_uds}
\ket{\psi_e^k}=\sum_{n=1}^k A_n^{(k)} \exp{(i\theta_n)}
\hat{a}_n^\dag\ket{0},
\end{equation}
with $A_n^{(k)}\in\mathbb{C}$ and where the normalization condition $\sum_{n=1}^k
\vert{}A_n^{(k)}\vert{}^2=1$ is always satisfied. 
The angles $\theta_n$ fulfill $\theta_n=0$ if the signal state identified by 
Eve's USD measurement at the time instance $n$ is $\ket{\alpha}$ and
$\theta_n=\pi$ if the signal state identified by Eve is $\ket{-\alpha}$, 
the operator $\hat{a}_n^\dag$ represents
a creation operator for one photon in temporal mode $n$, and 
the state $\ket{0}$
refers to the vacuum state. Eq.~(\ref{signal_uds}) considers the 
possibility of using different amplitudes $A_n^{(k)}$ for the 
resent signals, following the spirit of Ref.~\cite{tsurumaru_dps}. 
The superscript $k$ labeling the coefficients $A_n^{(k)}$  
emphasizes the fact that the value of these coefficients 
may depend on the number of temporal modes
$k$ contained in $\ket{\psi_e^k}$. 
Moreover, from now on we will
use the convention that the first temporal mode of $\ket{\psi_e^k}$
that arrives at Bob's detection device is mode $n=k$, while the last
one is mode $n=1$. This labeling convention is illustrated in 
Fig.~\ref{ordering}.

Let us now determine the minimum number, $1+d$, of vacuum states that Eve needs 
to send to Bob after each signal state $\ket{\psi_e^k}$.
From the previous 
paragraph we learn that whenever Bob receives a state $\ket{\psi_e^k}$ 
satisfying 
Eq.~(\ref{signal_uds}) then he obtains one single click in his detection device. 
This click can occur, however, in 
any temporal mode $n$, with $n\in[0,k]$ \cite{Note5}. The minimum value of 
the parameter $d$ can be calculated from the case where Bob obtains a click in 
the last possible temporal mode, {\it i.e.}, $n=0$ (see Fig.~\ref{ordering}).
Let us assume that such a click occurs, and let $t_d$ and $f_c$ 
denote, respectively, the dead-time of Bob's detectors and the clock 
frequency of the system. To guarantee that Bob cannot obtain any other 
click from a following signal state until $t_d$ finishes we 
find that $d$ has to fulfill
$d\geq\lceil{}t_df_c\rceil$.
That is, the parameter $d$ has to be larger than or equal to the number of signal
states sent by Alice within a period of time equal to the dead-time. From now on 
we shall consider that Eve selects $d$ such that 
\begin{equation}\label{eq_d}
d=\lceil{}t_df_c\rceil.
\end{equation}
 
Next, we obtain an expression for the gain and for the QBER
introduced by Eve in this scenario. 

\subsection{Gain}\label{gain_A}

The gain, that we shall denote as $G$, of a sequential attack is
defined as the probability that Bob obtains a click per signal
state sent by Alice. It can be expressed as $G=N_{clicks}/N$,
where $N_{clicks}$ represents the average total number of clicks
obtained by Bob, and $N$ is the total number of signal states
sent by Alice. The parameter $N_{clicks}$ can be expressed as
$N_{clicks}=(N/N^{e})N_{clicks}^{e}$, with $N^e$ denoting the
average total number of pulses of signal states sent by 
Eve (see Fig.~\ref{strategy1_meas}), and where
$N_{clicks}^{e}$ represents the average total number of clicks
obtained by Bob when Eve sends to him precisely these signal
states. With this notation, the gain of a sequential attack can
be written as
\begin{equation}\label{gain}
G=\frac{N_{clicks}^{e}}{N^{e}}.
\end{equation}

Next, we obtain an expression for $N_{clicks}^e$ and $N^{e}$. Let
us begin with $N_{clicks}^e$. Whenever Eve sends to Bob a signal
state $\ket{\psi_e^k}$ followed by $1+d$ vacuum states (Cases A, B,
and C in Fig.~\ref{strategy1_meas}) Bob always obtains one click
in his detection apparatus. On the other hand, if Eve sends to
Bob only vacuum states (Cases D and E in
Fig.~\ref{strategy1_meas}) Bob never obtains a click. This means, 
in particular,
that $N_{clicks}^e$ can be expressed as
\begin{equation}
N_{clicks}^e=\sum_{k=M_{min}}^{M_{max}} p_s(k),
\end{equation}
with $p_s(k)$ given by Eq.~(\ref{ps_k}). 
This expression can be further simplified as
\begin{equation}\label{N_clicks}
N_{clicks}^e=\big[q+(1-q)p\big]p^{M_{min}}.
\end{equation}

The analysis to obtain $N^{e}$ is similar. A signal state
$\ket{\psi_e^k}$ followed by $1+d$ vacuum states can be seen as
containing $k+1+d$ pulses. On the other hand, the number of vacuum
pulses alone that Eve sends to Bob can vary from $1$ to $M_{min}+1$
(see Fig.~\ref{strategy1_meas}).
Adding all these terms together, and taking into account their a
priori probabilities, we obtain that $N^{e}$ can be written as
\begin{equation}
N^e=\sum_{k=0}^{M_{max}} p_v(k)(k+1)+p_s(k)(k+1+d),
\end{equation}
with $p_v(k)$ given by Eq.~(\ref{pv_k}).
This expression can be simplified as
\begin{equation}\label{n_clicks_a}
N^e=\frac{d\big[q+(1-2q)p-(1-q)p^{2}\big]p^{M_{min}}
-p^{M_{max}+1}+1}{1-p}
\end{equation}

The gain $G$ can be related with a transmission distance $l$ for a given QKD scheme, 
{\it i.e.}, a distance which provides an expected click rate at Bob's side given 
by $G$. This last condition can be written as 
\begin{equation}\label{nuevag}
G=1-\exp{(-\mu_\alpha\eta_{det}\eta_t)},
\end{equation}
where $\eta_{det}$ represents the detection efficiency of the detectors employed by Bob,
and $\eta_t$ denotes the transmittivity of the quantum channel. In the 
case of a DPS QKD scheme, the value of $\eta_t$ can be derived from the loss 
coefficient $\gamma$ of the optical fiber measured in dB/km, the 
transmission distance $l$ measured in km, and the loss in Bob's 
interferometer $L$ measured in dB as
\begin{equation}\label{nuevag2}
\eta_t=10^{-\frac{\gamma{}l+L}{10}}.
\end{equation}
From Eq.~(\ref{nuevag}) and Eq.~(\ref{nuevag2}), we find that the transmission distance 
$l$ that provides a gain $G$ is given by
\begin{equation}
l=-\frac{1}{\gamma}\bigg[L+10\log_{10}\bigg(\frac{-\ln{(1-G)}}{\mu_\alpha\eta_{det}}\bigg)\bigg].
\end{equation} 

\subsection{Quantum bit error rate}\label{qber_A}

The QBER, that we shall denote as $Q$, is defined as
$Q=N_{errors}/N_{clicks}$, where $N_{errors}$ represents the
average total number of errors obtained by Bob, and $N_{clicks}$
is again the average total number of clicks at Bob's side. The
parameter $N_{errors}$ can be expressed as
$N_{errors}=(N/N^e)N_{errors}^e$, with $N_{errors}^e$ denoting the
average total number of errors obtained by Bob when Eve sends
him the different signal states considered in her strategy (see
Fig.~\ref{strategy1_meas}). With this notation, and using again
the fact that $N_{clicks}=(N/N^{e})N_{clicks}^{e}$, we obtain that
the QBER of a sequential attack can be expressed as
\begin{equation}\label{eqqber_g}
Q=\frac{N_{errors}^e}{N_{clicks}^e}.
\end{equation}
The parameter $N_{clicks}^e$ was calculated in the previous section 
and it is given by Eq.~(\ref{N_clicks}). 
Let us now obtain an expression for $N_{errors}^e$. We shall
distinguish the same cases like in the previous section, depending
on the type of signal states that Eve sends to Bob. Whenever Eve
sends to Bob a signal state $\ket{\psi_e^k}$ followed by $1+d$ vacuum
states (Cases A, B, and C in Fig.~\ref{strategy1_meas}) then 
we shall denote 
the average total number of errors in this scenario as $e(k)$.
On the other hand, if Eve sends to Bob only vacuum states (Cases
D and E in Fig.~\ref{strategy1_meas}) Bob never obtains an error.
This means, in particular, that $N_{errors}^e$ can be expressed as
\begin{equation}
N_{errors}^e=\sum_{k=M_{min}}^{M_{max}} p_s(k)e(k).
\end{equation}
The parameters $e(k)$, with $M_{min}\leq{}k\leq{}M_{max}$, can be
obtained from the signal states $\ket{\psi_e^k}$, together with the
detection device used by Bob. They are calculated in 
Appendix~\ref{ap_a} and are given by
\begin{equation}\label{eq_ek}
e(k)=\frac{1}{4}\Bigg[\vert{}A_1^{(k)}\vert{}^2+\sum_{n=1}^{k-1}
\vert{}A_{n+1}^{(k)}-A_n^{(k)}\vert{}^2+\vert{}A_k^{(k)}\vert{}^2\Bigg].
\end{equation}

\subsection{Evaluation}\label{primera_evalua}

We have seen above that a sequential attack can
be parametrized by the minimum 
number $M_{min}$ of consecutive USD successful results that Eve
needs to obtain in order to consider the sequence of measurement
outcomes successful, the maximum number $M_{max}$ of
consecutive successful results that Eve can obtain in order
to send to Bob a train of signal states, the value of the probability 
$q$, {\it i.e.}, the probability that Eve actually
decides to send to Bob the signal
state $\ket{\psi_e^{M_{min}}}$ followed by $1+d$ vacuum states instead
of $M_{min}+1$ vacuum states, and the 
state coefficients $A_n^{(k)}\in\mathbb{C}$ that characterize the signal 
states $\ket{\psi_e^{k}}$, with $M_{min}\leq{}k\leq{}M_{max}$.

\begin{figure}
\begin{center}
{\rotatebox{-90}{\includegraphics[scale=0.31]{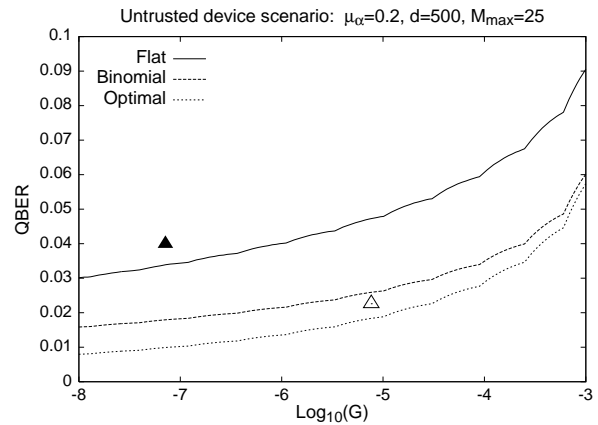}}}
\end{center}
\caption{Gain ($G$) versus QBER in a sequential attack for
three different distributions of the state coefficients $A_n^{(k)}$: 
flat (solid), binomial (dashed), and the optimal distribution
(dotted). The mean photon number of Alice's signal
states is $\mu_{\alpha}=0.2$, and the parameter $d=500$. 
The triangles represent experimental data from Ref.~\cite{dpsqkd_exp3}. \label{exp1011}}
\end{figure}

\begin{figure}
\begin{center}
{\rotatebox{-90}{\includegraphics[scale=0.31]{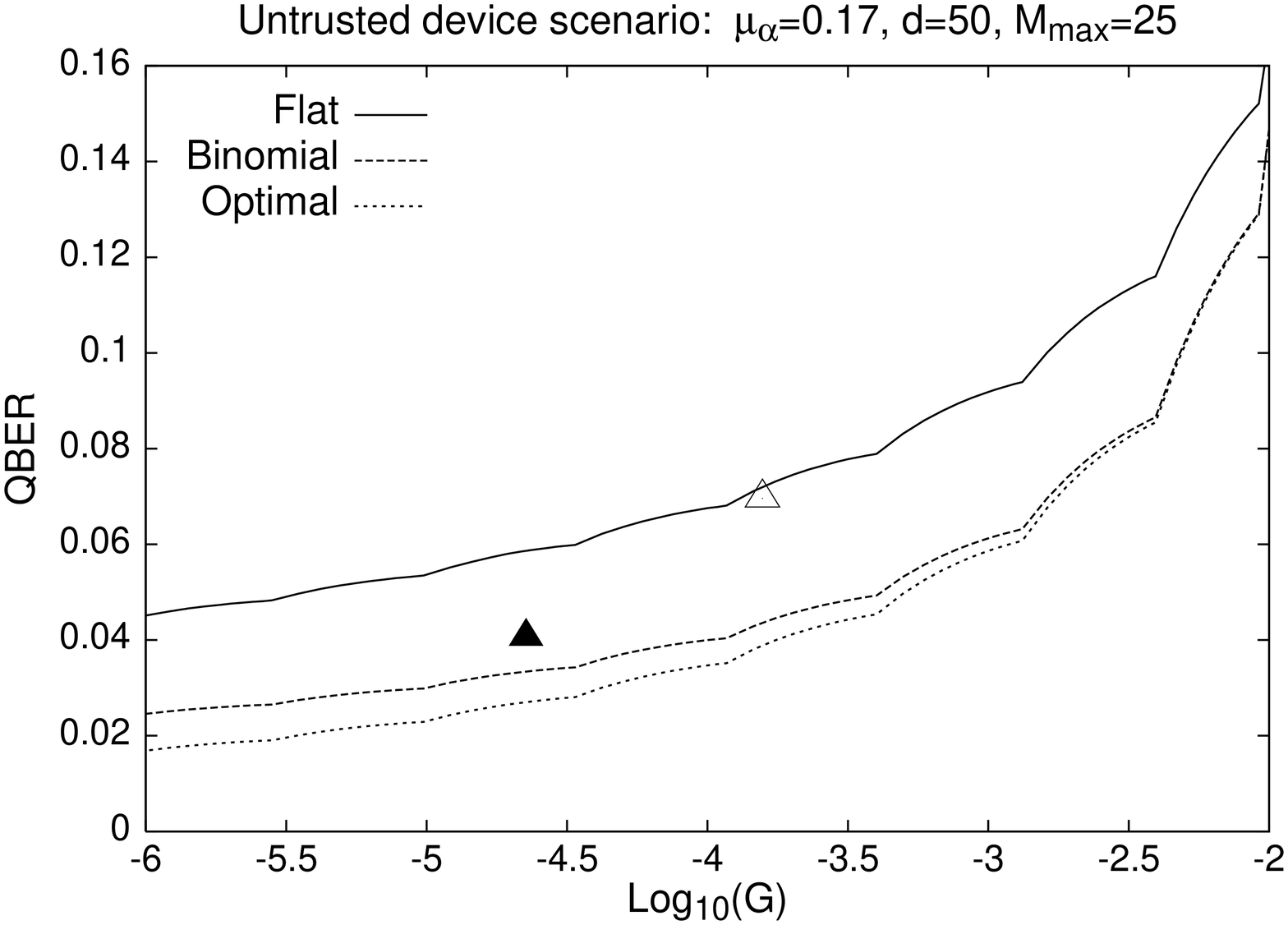}}}
\end{center}
\caption{Gain ($G$) versus QBER in a sequential attack for
three different distributions of the state coefficients $A_n^{(k)}$: 
flat (solid), binomial (dashed), and the optimal distribution
(dotted). The mean photon number of Alice's signal
states is $\mu_{\alpha}=0.17$, and the parameter $d=50$. 
The triangles represent experimental data from Ref.~\cite{dpsqkd_exp1}. 
(See also Ref.~\cite{dpsqkd_exp2b}.) \label{exp34}}
\end{figure}

\begin{figure}
\begin{center}
{\rotatebox{-90}{\includegraphics[scale=0.31]{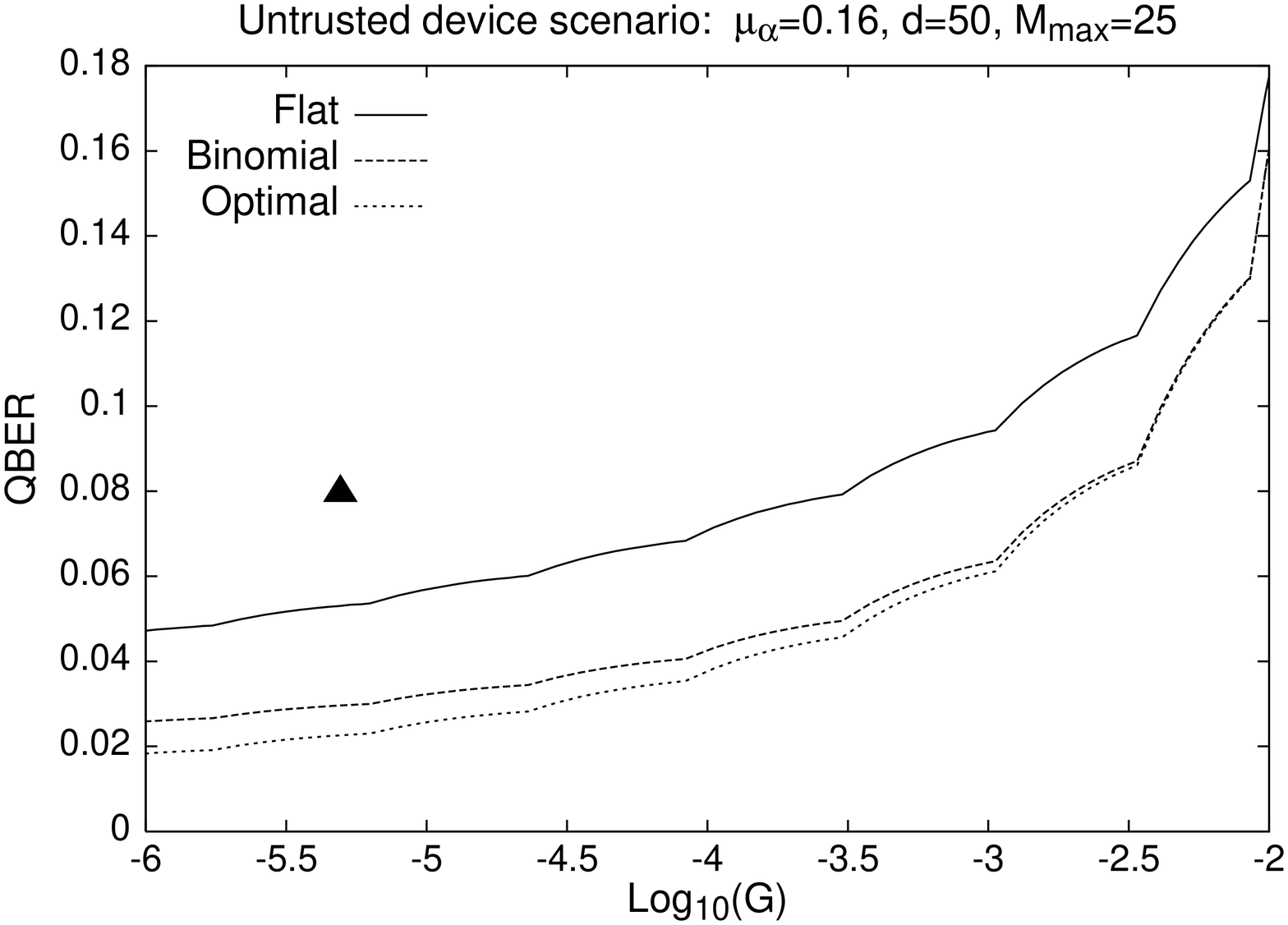}}}
\end{center}
\caption{Gain ($G$) versus QBER in a sequential attack for
three different distributions of the state coefficients $A_n^{(k)}$: 
flat (solid), binomial (dashed), and the optimal distribution
(dotted). The mean photon number of Alice's signal
states is $\mu_{\alpha}=0.16$, and the parameter $d=50$. 
The triangle represents experimental data from Ref.~\cite{dpsqkd_exp1}. 
(See also Ref.~\cite{dpsqkd_exp2b}.) \label{exp5}}
\end{figure}

\begin{figure}
\begin{center}
{\rotatebox{-90}{\includegraphics[scale=0.31]{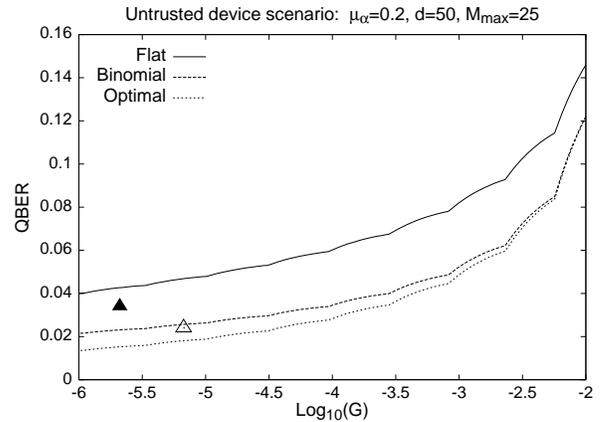}}}
\end{center}
\caption{Gain ($G$) versus QBER in a sequential attack for
three different distributions of the state coefficients $A_n^{(k)}$: 
flat (solid), binomial (dashed), and the optimal distribution
(dotted). The mean photon number of Alice's signal
states is $\mu_{\alpha}=0.2$, and the parameter $d=50$. 
The triangles represent experimental data from Ref.~\cite{dpsqkd_exp2}. 
(See also Ref.~\cite{dpsqkd_exp2b}.) \label{exp69}}
\end{figure}

Figures~\ref{exp1011}, \ref{exp34}, \ref{exp5} and \ref{exp69} show a 
graphical representation of the Gain
versus the QBER in this sequential attack for different values of the mean 
photon-number $\mu_{\alpha}$ of Alice's signal states, the 
parameter $d$, and the 
state coefficients $A_n^{(k)}$. It states that no key distillation protocol
can provide a secret key from the correlations established by the users 
above the curves, {\it i.e.}, the secret key rate in that region is 
zero. In these examples we 
consider three possible distributions for $A_n^{(k)}$: the flat
distribution, the binomial distribution, and we also calculate the 
optimal distribution, {\it i.e.}, the one which 
provides the lowest QBER for a given value of
the Gain. The corresponding state coefficients for these distributions are given by
\begin{eqnarray}\label{distribu}
\text{Flat:}&&A^{(k)}_n=\frac{1}{\sqrt{k}},\;\text{for all } n\in[1,k], \nonumber \\
\text{Binomial:} && A^{(k)}_n = \left(\frac{1}{\sqrt{2}}\right)^{k-1} \sqrt{\left( \begin{array}{c} k-1 \\ n-1 \end{array}\right)},
\end{eqnarray}
while the method to obtain the optimal distribution is described in Appendix~\ref{ap_opt}.
Figures~\ref{exp1011}, \ref{exp34}, \ref{exp5} and \ref{exp69}
assume as well that $M_{max}$ is fixed 
and given by $M_{max}=25$, and we vary the parameters $M_{min}<M_{max}$ and $q\in[0,1]$.  
They also include 
experimental data from Refs.~\cite{dpsqkd_exp1,dpsqkd_exp2,dpsqkd_exp2b,dpsqkd_exp3}.
For instance, in the experiment reported in Ref.~\cite{dpsqkd_exp3} the dead-time 
of Bob's detectors is $t_d=50$ ns and the clock frequency of the system is 
$f_c=10$ GHz. From Eq.~(\ref{eq_d}) we obtain, therefore, that 
$d=500$. (See Fig.~\ref{exp1011}.) Similarly, in the experiments realized in
Refs.~\cite{dpsqkd_exp1,dpsqkd_exp2,dpsqkd_exp2b} we have that $t_d=50$ ns and 
$f_c=1$ GHz. This means, in particular, that in all these cases $d=50$. 
(See Figs.~\ref{exp34}, \ref{exp5} and \ref{exp69}.)
According to our results it seems that 
all the long-distance 
implementations of DPS QKD reported in Refs.~\cite{dpsqkd_exp1,dpsqkd_exp2,dpsqkd_exp2b,dpsqkd_exp3}
would be insecure against a sequential attack in the untrusted 
device scenario. That is, there exists no improved classical
communication protocol or improved security analysis which
might allow the data of Refs.~\cite{dpsqkd_exp1,dpsqkd_exp2,dpsqkd_exp2b,dpsqkd_exp3} 
to be turned into secret key.

\section{Trusted device scenario}\label{medattack}

In this section we impose constraints on Eve's capabilities, and
we are interested in the effect that these constraints have on her
eavesdropping strategy. In particular, we study the situation
where Eve is not able to manipulate Alice and Bob's devices at
all, but she is limited to act exclusively on the quantum channel
(See, e.g., Refs.~\cite{jahma01,curty_pns}). That is, we shall
consider that the detection efficiency, the dark count
probability, and the dead-time of Bob's detectors are now fixed
by the actual experiment, and Eve cannot influence them to obtain
extra information about the shared key. The main motivation to
analyze this scenario is that, from a practical point of view, it
constitutes a reasonable description of a realistic situation,
where Alice and Bob could in principle limit Eve's influence on their apparatus
by some counterattack techniques \cite{Note6}. Moreover, 
this could only enhance Alice and Bob's ability to distill a 
secret key.

The detection efficiency $\eta_{det}$ of Bob's detectors  
typically satisfies $\eta_{det}<1$. Therefore, in this 
scenario, Eve might be interested 
in sending Bob multi-photon signals, instead 
of single-photon
states like in Sec.~\ref{susda}, in order to increase the gain. 
Moreover, as mentioned above, we assume 
now that the dead-time of Bob's detectors is already present in 
$D0$ and $D1$ and Eve does not 
need to select her signal states such that they can reproduce it.  
These two facts motivate the following definition for the signal
states $\rho_e^k$ in this case. In particular, we shall consider that $\rho_e^k$
consists of a classical mixture of pure states, that we 
shall denote as $\ket{\psi_k^m}$, containing $m\geq{}1$ photons that are
distributed among $k$ temporal modes, {\it i.e.}, 
\begin{equation}\label{signals_tds}
\rho_e^k=\sum_{m=1}^{\infty} p_m \ket{\psi_k^m}\bra{\psi_k^m},
\end{equation}
with the photon-number probabilities $p_m$ satisfying
$\sum_{m=1}^{\infty} p_m=1$ \cite{Note7}. The states $\ket{\psi_k^m}$ 
are
defined as
\begin{equation}\label{pure_signals_tds}
\ket{\psi_k^m}=\frac{\big(\hat{\psi}_{k,m}^{\dag}\big)^m}{\sqrt{m!}}
\ket{0},
\end{equation}
with $\ket{0}$ representing the vacuum state, and where the
operators $\hat{\psi}_{k,m}^{\dag}$ are given by
\begin{equation}\label{operator_tds}
\hat{\psi}_{k,m}^{\dag}=\sum_{n=1}^k A_{n,m}^{(k)} \exp{(i\theta_n)}
\hat{a}_n^\dag.
\end{equation}
As before, $\hat{a}_n^\dag$ represents a creation
operator for one photon in temporal mode $n$, and the
coefficients $A_{n,m}^{(k)}$ satisfy the normalization condition
$\sum_{n=1}^k \vert{}A_{n,m}^{(k)}\vert{}^2=1$. The superscript $k$
and the subscript $m$ labeling these coefficients are used to 
emphasize that the value of
$A_{n,m}^{(k)}$ may depend, respectively, on the number of
temporal modes $k$, and on the number of photons $m$, contained in
$\ket{\psi_k^m}$. Moreover, like in Sec.~\ref{susda}, 
we shall consider that the time
difference between two consecutive temporal modes in
$\ket{\psi_k^m}$ is set equal to the time difference $\Delta{}t$
between two consecutive pulses sent by Alice. The definition of
$\theta_n$ in Eq.~(\ref{operator_tds}) is also equal to the one
provided for these angles in Sec.~\ref{susda}. That is, 
$\theta_n=0$ if the signal state identified by Eve's
USD measurement at the time instance $n$ is $\ket{\alpha}$, and
$\theta_n=\pi$ if the state identified by Eve is $\ket{-\alpha}$. 
Besides, since Eve does not need to choose an
eavesdropping strategy that reproduces Bob's detectors dead-time, 
the number of vacuum states that she sends to him
following each signal state $\rho_e^k$ can be set equal to one,
{\it i.e.}, we will assume that $d=0$ in
Fig.~\ref{strategy1_meas}. This vacuum state corresponds to the 
inconclusive result. Of course, Eve could choose as well an
eavesdropping strategy where the 
parameter $d$ satisfies $d>0$; this strategy would only 
cause that the value of the gain decreases and, therefore,
it would also diminish the strength of Eve's attack. 
Finally, for simplicity, we shall consider that 
the parameter $M_{max}$ 
satisfies $M_{max}\leq{}\lceil{}t_df_c\rceil$. 
This condition guarantees that, within each of the blocks of 
signal states illustrated in Fig.~\ref{strategy1_meas}, 
Bob can obtain, at most, only {\it one} click in his detection
apparatus.

Next, we obtain an expression for the gain and for the QBER
introduced by Eve in this scenario. We will analyze as well 
the resulting
double click rate at Bob's side in this eavesdropping
strategy. Note that now the double click rate obtained by Bob may 
increase due to the multi-photon signals used by Eve.

\subsection{Gain}\label{gain_A_tds}

As shown previously, in this sequential attack the gain is given
by Eq.~(\ref{gain}). However, the analysis to obtain an
expression for the parameters $N_{clicks}^{e}$ and $N^{e}$ is now
slightly different from the one considered in Sec.~\ref{susda},
where Bob's detectors dead-time was reproduced by the signal
states sent by Eve. In particular, now we need to include the
effect of the dead-time of Bob's detectors in the detection model.
Moreover, in this scenario Bob can obtain as well double clicks
in his detection apparatus. We shall consider that these double click events 
are not discarded by Bob, but they contribute to the raw 
key. Every time Bob obtains a
double click, he just decides randomly the bit value \cite{Norbert99}.

\begin{figure}
\begin{center}
\includegraphics[scale=1.1]{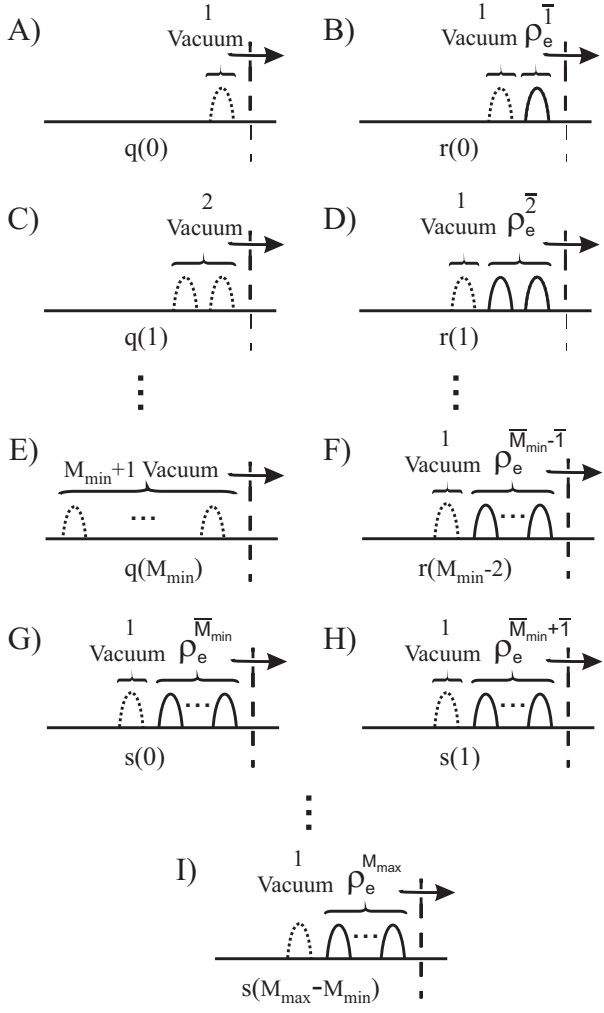}
\end{center}
\caption{Possible signal states arriving at Bob's detection
apparatus after a dead-time, together with their a
priori probabilities. The arrow indicates the transmission
direction. \label{strategy1_meas_dead_time}}
\end{figure}

Let us start by considering again the type of signal states that 
Eve sends to Bob in this strategy. These signals are illustrated in
Fig.~\ref{strategy1_meas}, where the
states $\rho_e^k$ are given by Eq.~(\ref{signals_tds}) and the parameter 
$d=0$. However, once Bob's detectors are recovered from 
a dead-time produced by a previous click, 
the first temporal
mode that arrives at Bob's side at this time instant might not
coincide with the first temporal mode of any of the blocks of
signal states considered in Fig.~\ref{strategy1_meas}. This 
first mode could be, in principle, any of the temporal modes
contained in these blocks of signal states. For instance, 
it could be any of the $k$ temporal modes contained in $\rho_e^k$. 
Note that this last case was never possible in the scenario
analyzed in Sec.~\ref{susda}. Figure~\ref{strategy1_meas_dead_time}
shows a graphical representation of the possible blocks of signal states that can arrive at
Bob's side after a dead-time. These blocks of signals are just
obtained from those illustrated in
Fig.~\ref{strategy1_meas} by discarding some of their first
temporal modes. For example, the block which contains only one vacuum 
state (Case A in Fig.~\ref{strategy1_meas_dead_time}) could arise 
from every block of signals considered in Fig.~\ref{strategy1_meas},
just by discarding all their temporal modes except the last one. The 
block composed by two vacuum states (Case C in Fig.~\ref{strategy1_meas_dead_time})
could originate from any block of signals in Fig.~\ref{strategy1_meas} 
that contains at least two vacuum states at the end of the block 
(Cases D and E in Fig.~\ref{strategy1_meas} when $d=0$), 
and so on. The a priori probabilities of the different blocks of signals illustrated in 
Fig.~\ref{strategy1_meas_dead_time}, that we shall
denoted as $q(k)$, $r(k)$, and $s(k)$, respectively, are
calculated in Appendix~\ref{ap_b}.
The superscript $\bar{k}$, with $1\leq{}\bar{k}<M_{max}$, that labels 
the states $\rho_e^{\bar{k}}$ in
Fig.~\ref{strategy1_meas_dead_time} is used to emphasize the fact that
these signals may correspond to the last $\bar{k}$ temporal modes of any
signal state $\rho_e^k$ with $k\geq{}\bar{k}$.

The parameters $N_{clicks}^{e}$ and $N^{e}$ can now be expressed,
respectively, as
\begin{equation}\label{n_e_clicks_a_tds}
N_{clicks}^{e}=\sum_{m=1}^{\infty} p_m N_{clicks}^{e}(m),
\end{equation}
where $N_{clicks}^{e}(m)$ denotes the average total number of
clicks obtained by Bob when he receives the different
blocks of signal states illustrated in
Fig.~\ref{strategy1_meas_dead_time} with $\rho_e^{\bar{k}}$ being
the last $\bar{k}$ temporal modes of any state $\ket{\psi_{k}^m}$,
and
\begin{equation}\label{n_e_a_tds}
N^{e}=\sum_{m=1}^{\infty} p_m N^{e}(m),
\end{equation}
with $N^{e}(m)$ representing the average total number of pulses
{\it caused} by these blocks of signal states. That is, 
to calculate $N^{e}(m)$ we need to consider not only the 
number of temporal modes contained in the signals illustrated 
in Fig.~\ref{strategy1_meas_dead_time} but also 
the fact that whenever Bob
obtains a click in any of these modes 
then he cannot obtain any other click during the
following $\tilde{d}=\lceil{}t_df_c\rceil$ pulses \cite{nsqp}.
The main idea behind the whole analysis contained in this 
section is to study the behaviour of the possible blocks of signals  
that Bob can receive from Eve after a dead-time, together 
with their a priori probabilities. 
Every time a dead-time
finishes we have a new trial of this random process.

Next, we calculate an expression for $N_{clicks}^e(m)$ and
$N^{e}(m)$. We shall distinguish several cases, depending on the
block of signal states that arrives at Bob's side after a dead-time.
Let us begin with Case A in Fig.~\ref{strategy1_meas_dead_time}.
The probability that Bob obtains a click in this scenario 
depends on the identity of the preceding signal. We shall
denote with $p_{vv}$ the probability to obtain a click when the
previous signal is also a vacuum state, and we shall denote with
$p_{vk}^m$ the probability to obtain a click when the previous
signal is the state $\ket{\psi_{k}^m}$. These two probabilities 
are calculated in Appendix~\ref{ap_c}.
In general, we have that 
$p_{vv}>0$ due
to the dark counts in Bob's detectors. Let $p_{pv}$ ($p_{pk}$) denote 
the probability that the previous signal is a vacuum state (the signal
$\ket{\psi_{k}^m}$). These probabilities are calculated in 
Appendix~\ref{ap_ppv}. With this notation, we find that the 
average total number of clicks in
this scenario, that we shall represent as $N_{clicks,q(0)}^e(m)$, is given by
\begin{equation}
N_{clicks,q(0)}^e(m)=p_{pv}p_{vv}+\sum_{k=M_{min}}^{M_{max}}
p_{pk}p_{vk}^m.
\end{equation}
Next, we calculate an expression for the average total number of pulses,
that we shall denote as $N^{e}_{q(0)}(m)$. 
This parameter depends on whether Bob's detectors click or do not click.
In particular, we have that whenever Bob obtains a click in his detection 
apparatus then the total
number of pulses that we need to consider is $1+\tilde{d}$. That is, 
in this case we need to include the
effect of the dead-time. Otherwise, the number of pulses is one.
We obtain, therefore, that
\begin{eqnarray}
N^{e}_{q(0)}(m)&=&p_{pv}[p_{vv}(1+\tilde{d})+(1-p_{vv})] \nonumber \\
\ &+&\sum_{k=M_{min}}^{M_{max}} p_{pk}[p_{vk}^m(1+\tilde{d})+(1-p_{vk}^m)] \nonumber \\
\ &=&p_{pv}(1+\tilde{d}p_{vv})+\sum_{k=M_{min}}^{M_{max}}
p_{pk}(1+\tilde{d}p_{vk}^m).\nonumber\\
\end{eqnarray}
The analysis of the remaining cases included in
Fig.~\ref{strategy1_meas_dead_time} is similar. Whenever Bob
receives a block of $k+1$ vacuum states (Cases C and E in
Fig.~\ref{strategy1_meas_dead_time}), with $1\leq{}k\leq{}M_{min}$, it
is guaranteed that the signal which precedes the block is always a vacuum state.
This is justified by the particular structure of the different blocks of signal
states that Eve can send to Bob (see Fig.~\ref{strategy1_meas}). In this case, 
due to the dead-time of Bob's detectors, 
Bob can only obtain a click in a given temporal mode 
if the previous modes of the block did not click. 
The average total number of clicks, that
we shall denote as $N_{clicks,q(k)}^e(m)$, can then be expressed as
\begin{equation}
N_{clicks,q(k)}^e(m)=\sum_{n=0}^k (1-p_{vv})^n{}p_{vv}=1-(1-p_{vv})^{k+1}.
\end{equation}
In order to calculate the average total number of pulses, that we shall 
denote as 
$N^{e}_{q(k)}(m)$, note that, like before, whenever Bob obtains a click in 
temporal mode $l\in[0,k]$ then the total
number of pulses that we need to consider is $(k-l)+(1+\tilde{d})$. 
The first term in the summation, $(k-l)$, represents the total number 
of modes contained in the block before the mode that actually 
clicked (see the labeling 
convention illustrated in Fig.~\ref{ordering}), while the second term, 
$(1+\tilde{d})$, includes the effect of the dead-time. Otherwise, 
the number of pulses is $k+1$. We find,
therefore, that $N^{e}_{q(k)}(m)$ can be written as
\begin{eqnarray}
N^{e}_{q(k)}(m)&=&(k+1)(1-p_{vv})^{k+1} \nonumber \\
\ &+&\sum_{n=0}^k
(n+1+\tilde{d})(1-p_{vv})^np_{vv} \nonumber \\
\ &=&\frac{(1+\tilde{d}p_{vv})\big[1-(1-p_{vv})^{k+1}\big]}{p_{vv}}.
\end{eqnarray}
When Bob receives a state $\rho_e^{\bar{k}}$, with
$1\leq{}\bar{k}\leq{}M_{min}-1$, followed by one vacuum state (Cases
B, D, and F in Fig.~\ref{strategy1_meas_dead_time}), then the state 
which precedes that block of signals can never be a vacuum state. 
Let $p_{\bar{k}}^k$ denote the probability that $\rho_e^{\bar{k}}$ 
stems from the last $\bar{k}$
temporal modes of a signal state $\ket{\psi_{k}^m}$ with 
$M_{min}\leq{}k\leq{}M_{max}$, and let $q_{\bar{k}}^m(k)$ be the average 
total number of clicks obtained by Bob in this scenario. 
The probability $p_{\bar{k}}^k$ is calculated in 
Appendix~\ref{ap_pkkb}, while the parameter $q_{\bar{k}}^m(k)$ can be written as
\begin{equation}
q_{\bar{k}}^m(k)=\sum_{n=0}^{\bar{k}} p_{c,\bar{k},k}^m(n),
\end{equation}
with $p_{c,\bar{k},k}^m(n)$ denoting the probability that Bob obtains a 
click in temporal mode $n$ and he does not 
obtain a click in any previous mode $l$ with $n<l\leq{}\bar{k}$. 
This last quantity is calculated 
in Appendix~\ref{ap_c}.
%
With this notation, we have that the average total number
of clicks in this case, that we shall denote as $N_{clicks,r(\bar{k}-1)}^e(m)$,
can be expressed as
\begin{equation}
N_{clicks,r(\bar{k}-1)}^e(m)=\sum_{k=M_{min}}^{M_{max}}
p_{\bar{k}}^k q_{\bar{k}}^m(k),
\end{equation}
  
Similarly, the average total number of pulses has now the form
\begin{equation}
N^{e}_{r(\bar{k}-1)}(m)=\sum_{k=M_{min}}^{M_{max}} p_{\bar{k}}^k
n_{\bar{k}}^m(k),
\end{equation}
where the parameter $n_{\bar{k}}^m(k)$ represents the average total 
number of pulses when Bob receives 
the last $\bar{k}$ temporal modes of the signal $\ket{\psi_{k}^m}$ together
with one vacuum state. This parameter is given by
\begin{eqnarray}
n_{\bar{k}}^m(k)&=&\sum_{n=0}^{\bar{k}} p_{c,\bar{k},k}^m(n)[\bar{k}-n+1+\tilde{d}] \nonumber \\
\ &+&[1-q_{\bar{k}}^m(k)](1+\bar{k}).
\end{eqnarray}

Bob can receive as well a state $\rho_e^{\bar{k}}$, with
$M_{min}\leq{}\bar{k}\leq{}M_{max}$, followed by one vacuum
state (Cases G, H and I in Fig.~\ref{strategy1_meas_dead_time}).
Let $p_{pv\bar{k}}$ denote the probability that the preceding 
signal is a vacuum state, and let $p_{p\bar{k}}^k$ 
be the probability that $\rho_e^{\bar{k}}$
stems from the last $\bar{k}$ temporal modes of the signal state
$\ket{\psi_{k}^m}$, with
$\bar{k}<k\leq{}M_{max}$. These two probabilities are calculated 
in Appendix~\ref{nac}. Using this notation, we obtain that 
the average total number of
clicks in this scenario, that we shall denote as
$N_{clicks,s(\bar{k}-M_{min})}^e(m)$, with
$M_{min}\leq{}\bar{k}\leq{}M_{max}$, can be expressed as
\begin{equation}\label{qnb}
N_{clicks,s(\bar{k}-M_{min})}^e(m)=p_{pv\bar{k}}q_{\bar{k}}^m(\bar{k})+\sum_{k=\bar{k}+1}^{M_{max}}
p_{p\bar{k}}^kq_{\bar{k}}^m(k),
\end{equation}
while the average total number of pulses, that we shall denote as
$N_{s(\bar{k}-M_{min})}^e(m)$, can directly be obtained from Eq.~(\ref{qnb}) 
just by substituting the parameters $q_{\bar{k}}^m(k)$, with 
$\bar{k}\leq{}k\leq{}M_{max}$, by $n_{\bar{k}}^m(k)$. 

Finally, $N_{clicks}^e(m)$ and $N^{e}(m)$ can be
calculated by adding all these terms together with their a
priori probabilities. That is,
\begin{eqnarray}\label{n_e_clicks_final}
N_{clicks}^e(m)&=&\sum_{k=0}^{M_{min}} q(k) N_{clicks,q(k)}^e(m) \nonumber\\
\ &+&\sum_{k=0}^{M_{min}-2} r(k) N_{clicks,r(k)}^e(m) \nonumber\\
\ &+&\sum_{k=0}^{M_{max}-M_{min}} s(k) N_{clicks,s(k)}^e(m),
\end{eqnarray}
and similarly for $N^{e}(m)$.

\subsection{Quantum bit error rate}\label{qber_a_tds}

The QBER in this strategy is given by Eq.~(\ref{eqqber_g}), with
the parameter $N_{clicks}^{e}$ now given by
Eq.~(\ref{n_e_clicks_a_tds}). In order to obtain $N_{errors}^e$
we follow the same method like in the previous section. In
particular, this quantity can now be expressed as
\begin{equation}
N_{errors}^{e}=\sum_{m=1}^{\infty} p_m N_{errors}^{e}(m),
\end{equation}
where $N_{errors}^{e}(m)$ denotes the average total number of
errors obtained by Bob when he receives from Eve the different
blocks of signal states illustrated in
Fig.~\ref{strategy1_meas_dead_time} with $\rho_e^{\bar{k}}$
representing the last $\bar{k}$ temporal modes of any state
$\ket{\psi_{k}^m}$ with $k\geq{}\bar{k}$. Note, moreover, that in
this scenario vacuum states can also produce errors in Bob's
detection apparatus due to the dark counts.

The analysis to obtain the parameter $N_{errors}^{e}(m)$ is completely
equivalent to the one included in Sec.~\ref{gain_A_tds} to
calculate $N_{clicks}^{e}(m)$; basically one only needs to substitute 
Bob's probabilities to obtain a
click $p_{vv}$, $p_{vk}^m$, and $p_{c,\bar{k},k}^m(n)$ by the 
probabilities to obtain an error
in precisely the same situations when these
probabilities were introduced. For instance, we need to
substitute $p_{vv}$ by the probability that Bob obtains an error
when he receives from Eve a vacuum state and the preceding signal
is also a vacuum state, and similar for the other cases. 
These
error probabilities, that we shall denote, respectively, as
$e_{vv}$, $e_{vk}^m$, and $p_{e,\bar{k},k}^m(n)$
are calculated in Appendix~\ref{ap_d}.
The only 
exception is the parameter $N_{errors,q(k)}^e(m)$ (Cases C and E in
Fig.~\ref{strategy1_meas_dead_time}), with $1\leq{}k\leq{}M_{min}$.
This exception only arises due to the notation used in Sec.~\ref{gain_A_tds}. 
Bob can obtain an error in a given temporal mode $n+1$ if the 
previous modes of the block did not {\it clicked}; that is, the 
probability to have an error in that mode is $(1-p_{vv})^n{}e_{vv}$. 
We obtain, therefore, that $N_{errors,q(k)}^e(m)$ is now given by
\begin{eqnarray}\label{cansadotarde}
N_{errors,q(k)}^e(m)&=&\sum_{n=0}^k (1-p_{vv})^n{}e_{vv} \nonumber \\
\ &=&e_{vv}+e_{vv}\frac{1-p_{vv}}{p_{vv}}\big[1-(1-p_{vv})^k\big]. \nonumber \\
\end{eqnarray}

\subsection{Double click rate}\label{dc_a_tds}

So far we have considered the case of the standard DPS QKD
protocol, where only the raw bit rate, together with the time
slots in which Bob obtains a click, are monitored. In this
section, however, we briefly analyze the case of an extended
version of the protocol, where Alice and Bob can also make use of
the double click rate at Bob's side to try to detect Eve.
That is, every time Bob obtains a double click in his 
detection apparatus he first records  
this event and, afterwards, he selects randomly the bit value. 
This is motivated by the fact
that, unlike the type of signal states considered in Sec.~\ref{susda}, 
now the states $\rho_e^k$ given by
Eq.~(\ref{signals_tds}) always present a non vanishing probability
of producing a double click. This means, in particular, that
Alice and Bob could employ this information to discard
those sequential attacks that increase the double click rate that 
they expect due to
the statistical fluctuations in the channel, together with the
effect of dark counts in Bob's detectors \cite{dcr_uds}.

The double click rate at Bob's side, that we shall denote as
$D_c$, is typically defined as $D_c=N_{D_c}/N$, where $N_{D_c}$
refers to the average total number of double clicks obtained by
Bob, and $N$ is again the total number of signal states sent by
Alice. $N_{D_c}$ can be expressed as $N_{D_c}=(N/N^e)N_{D_c}^e$,
with $N^e$ given by Eq.~(\ref{n_e_a_tds}) and where $N_{D_c}^e$
denotes the average total number of double clicks obtained by Bob
when he receives from Eve the different blocks of signals
illustrated in Fig.~\ref{strategy1_meas_dead_time}. With this
notation, we find that $D_c$ can be written as
\begin{equation}
D_c=\frac{N_{D_c}^e}{N^e}.
\end{equation}
The parameter $N_{D_c}^e$ can be expressed as
\begin{equation}\label{n_e_dc_a_tds}
N_{D_c}^e=\sum_{m=1}^{\infty} p_m N_{D_c}^e(m),
\end{equation}
where $N_{D_c}^e(m)$ denotes the average total number of double
clicks obtained by Bob when the signal states $\rho_e^{\bar{k}}$
illustrated
in Fig.~\ref{strategy1_meas_dead_time} represent the last 
$\bar{k}$ temporal modes of any state
$\ket{\psi_{k}^m}$.

Again, the analysis to obtain
$N_{D_c}^{e}(m)$ is completely equivalent to the one included in
Sec.~\ref{gain_A_tds} to calculate $N_{clicks}^{e}(m)$. We only
need to substitute in Eq.~(\ref{n_e_clicks_final}) the
probabilities to obtain a click $p_{vv}$, $p_{vk}^m$, and 
$p_{c,\bar{k},k}^m(n)$ by the probabilities to obtain a
double click in the same situation. We shall denote 
these double click probabilities as $dc_{vv}$, $dc_{vk}^m$ and 
$p_{dc,\bar{k},k}^m(n)$, and they are calculated in Appendix~\ref{ap_e}. The only 
exception is the parameter $N_{D_c,q(k)}^e(m)$ (Cases C and E in
Fig.~\ref{strategy1_meas_dead_time}), with $1\leq{}k\leq{}M_{min}$.
The reason for this exception is similar to the one presented 
in Sec.~\ref{qber_a_tds} for the parameter $N_{errors,q(k)}^e(m)$.  
In particular, $N_{D_c,q(k)}^e(m)$ can be obtained from Eq.~(\ref{cansadotarde}) by substituting 
the probability $e_{vv}$ by $dc_{vv}$. 

\subsection{Evaluation}\label{evalua2}

\begin{figure}
\begin{center}
{\rotatebox{-90}{\includegraphics[scale=0.31]{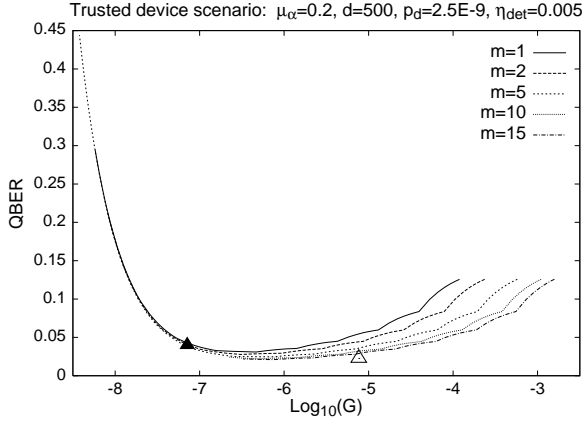}}}
\end{center}
\caption{Gain ($G$) versus QBER in a sequential attack for
different values of the photon number $m$, and for the 
optimal distribution of the state coefficients $A_{n,m}^{(k)}$
derived in Sec.~\ref{susda}. The mean photon number of Alice's signal
states is $\mu_{\alpha}=0.2$, the parameter $\tilde{d}=500$, the dark count 
probability of Bob's detectors is $p_d=2.5\times{}10^{-9}$, and the
detection efficiency $\eta_{det}=0.005$.
The triangles represent experimental data from Ref.~\cite{dpsqkd_exp3}. \label{exp1011_d_50t_b}}
\end{figure}

\begin{figure}
\begin{center}
{\rotatebox{-90}{\includegraphics[scale=0.31]{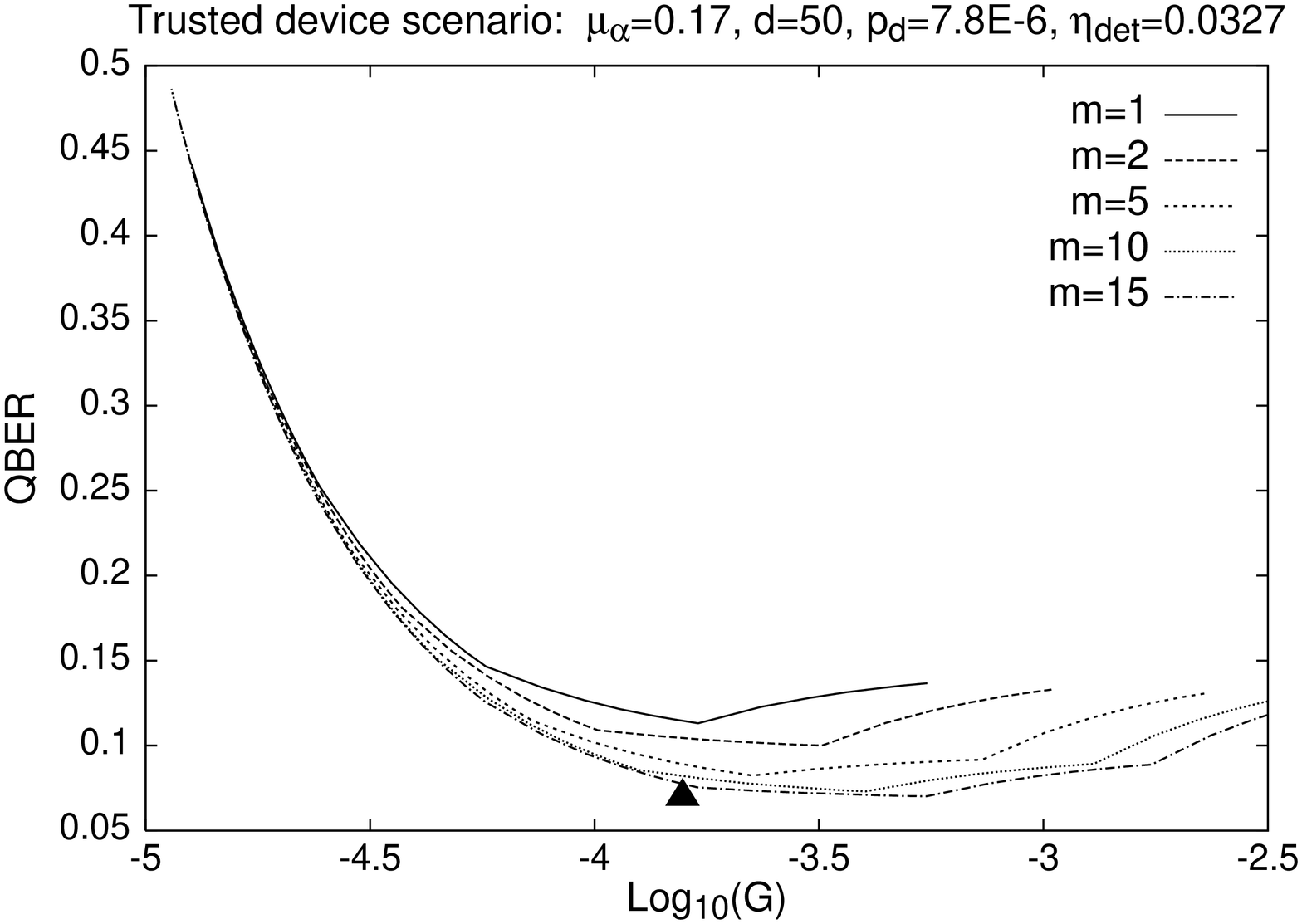}}}
\end{center}
\caption{Gain ($G$) versus QBER in a sequential attack for
different values of the photon number $m$, and for the 
optimal distribution of the state coefficients $A_{n,m}^{(k)}$
derived in Sec.~\ref{susda}. The mean photon number of Alice's signal
states is $\mu_{\alpha}=0.17$, the parameter $\tilde{d}=50$, the dark count 
probability of Bob's detectors is $p_d=7.8\times{}10^{-6}$, and the
detection efficiency $\eta_{det}=0.0327$.
The triangle represents experimental data from Ref.~\cite{dpsqkd_exp1}. 
(See also Ref.~\cite{dpsqkd_exp2b}.) \label{exp3tb}}
\end{figure}

\begin{figure}
\begin{center}
{\rotatebox{-90}{\includegraphics[scale=0.31]{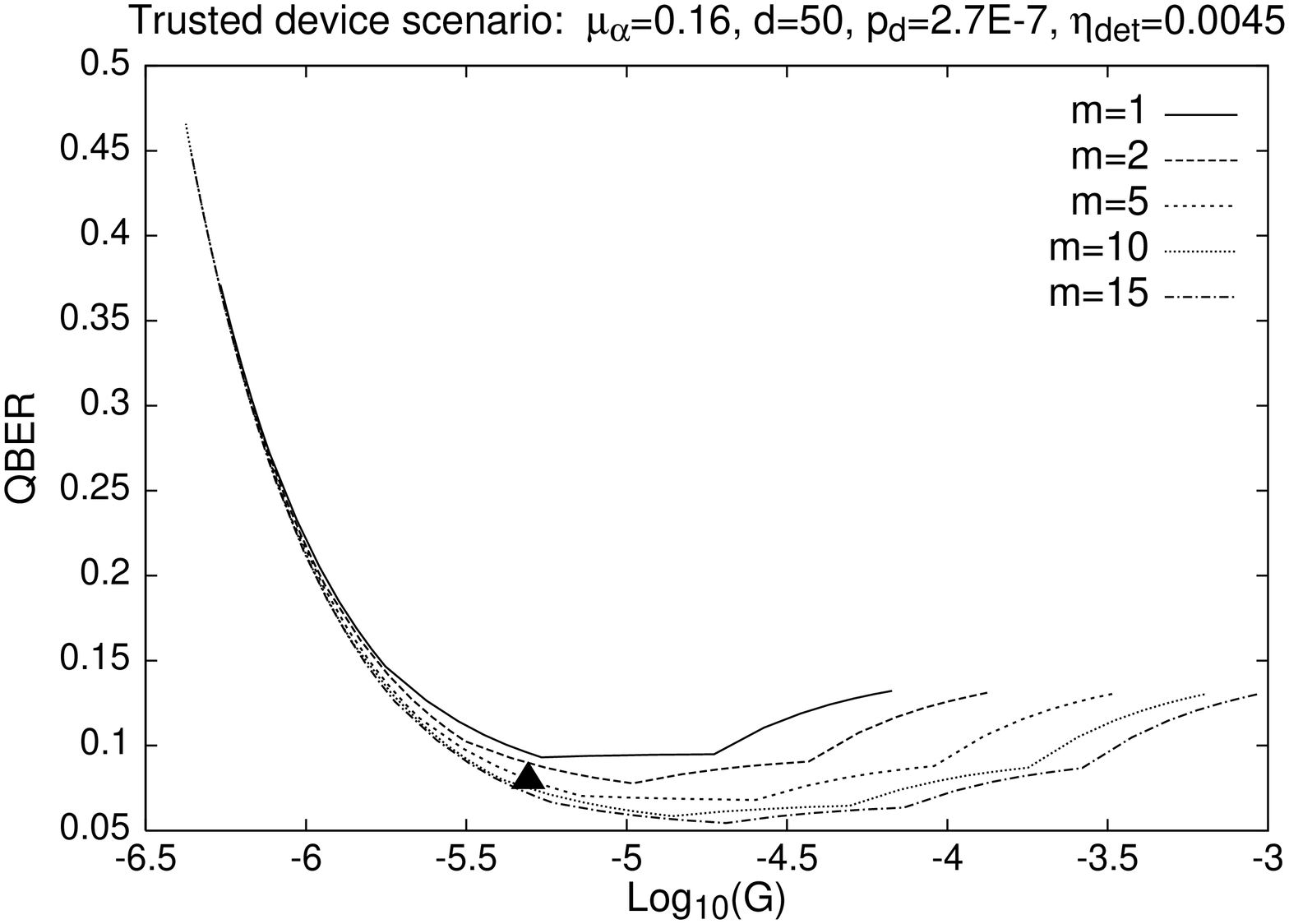}}}
\end{center}
\caption{Gain ($G$) versus QBER in a sequential attack for
different values of the photon number $m$, and for the 
optimal distribution of the state coefficients $A_{n,m}^{(k)}$
derived in Sec.~\ref{susda}. The mean photon number of Alice's signal
states is $\mu_{\alpha}=0.16$, the parameter $\tilde{d}=50$, the dark count 
probability of Bob's detectors is $p_d=2.7\times{}10^{-7}$, and the
detection efficiency $\eta_{det}=0.0045$.
The triangle represents experimental data from Ref.~\cite{dpsqkd_exp1}. 
(See also Ref.~\cite{dpsqkd_exp2b}.) \label{exp5tb}}
\end{figure}

\begin{figure}
\begin{center}
{\rotatebox{-90}{\includegraphics[scale=0.31]{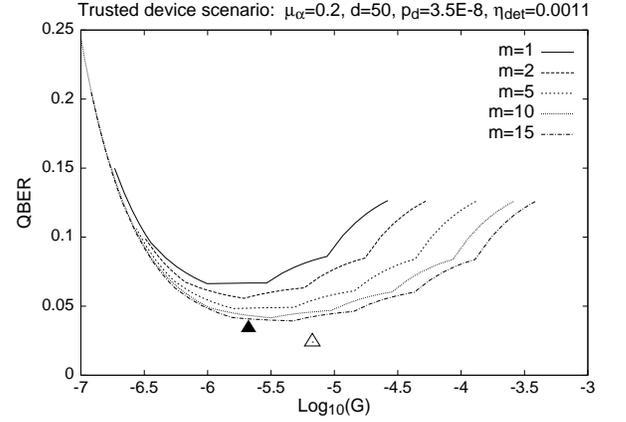}}}
\end{center}
\caption{Gain ($G$) versus QBER in a sequential attack for
different values of the photon number $m$, and for the 
optimal distribution of the state coefficients $A_{n,m}^{(k)}$
derived in Sec.~\ref{susda}. The mean photon number of Alice's signal
states is $\mu_{\alpha}=0.2$, the parameter $\tilde{d}=50$, the dark count 
probability of Bob's detectors is $p_d=3.5\times{}10^{-8}$, and the
detection efficiency $\eta_{det}=0.0011$.
The triangles represent experimental data from Ref.~\cite{dpsqkd_exp2}. 
(See also Ref.~\cite{dpsqkd_exp2b}.) \label{exp79tb}}
\end{figure}

In Figs.~\ref{exp1011_d_50t_b}, \ref{exp3tb}, \ref{exp5tb}  
and \ref{exp79tb} we plot the gain $G$ versus the QBER in a sequential attack for 
different values of the mean photon number $\mu_{\alpha}$
of Alice's signal states, the parameter $\tilde{d}$, the dark count 
probability $p_d$ and the detection efficiency $\eta_{det}$
of Bob's detectors, and the photon number $m$ of Eve's signal states \cite{dete_effi}. 
These examples illustrate the case of the standard DPS QKD protocol where Alice and Bob do 
not monitor separately the double click rate at Bob's side and Eve can select the 
parameter $m$ without any restriction
on the maximum tolerable double click rate. As before, these figures state that the secret 
key rate above the curves is zero. 

We 
fix the value of $M_{max}=\tilde{d}$ and we vary the value of the parameters $M_{min}<M_{max}$
and $q\in[0,1]$ like in Sec.~\ref{primera_evalua}. 
Moreover, for simplicity, we select the state coefficients 
$A_{n,m}^{(k)}$ of the signal states $\rho^k_{e}$ given by Eq.~(\ref{signals_tds}) 
as $A_{n,m}^{(k)}=A_{n}^{(k)}$ for all $m\geq{}1$, 
with $A_{n}^{(k)}$ given by the optimal distribution derived 
in Appendix~\ref{ap_opt} for the case of the untrusted 
device scenario. It can be shown that also in this case this 
distribution provides 
a lower QBER than the one obtained with a flat or a binomial distribution.  
As expected, the QBER produced by a sequential attack starts decreasing
as the losses in the channel increase, and, at some
point, it begins to increase again. This inflexion point is due to the 
dark count probability of Bob's detectors, which Eve cannot manipulate
in the trusted device scenario. In particular, 
when the gain is low enough such that most of the clicks obtained by Bob originate
from the dark counts of his detectors then the QBER starts increasing again. 
In the limit case where all the detected events arise from dark counts we
have that the QBER=0.5. Figs.~\ref{exp1011_d_50t_b}, \ref{exp3tb}, \ref{exp5tb}  
and \ref{exp79tb} also include experimental data from 
Refs.~\cite{dpsqkd_exp1,dpsqkd_exp2,dpsqkd_exp2b,dpsqkd_exp3}.
According to our results, and for the values of the parameter $m$ considered
in these examples, we find that the $200$ km DPS QKD experiment reported in 
Ref.~\cite{dpsqkd_exp3} together with the $105$ km DPS QKD experiment
reported in Ref.~\cite{dpsqkd_exp1}
would be insecure against a sequential attack even in the trusted 
device scenario. That is, the data of these experiments could never
be turned into secret key.

\begin{figure}
\begin{center}
{\rotatebox{-90}{\includegraphics[scale=0.31]{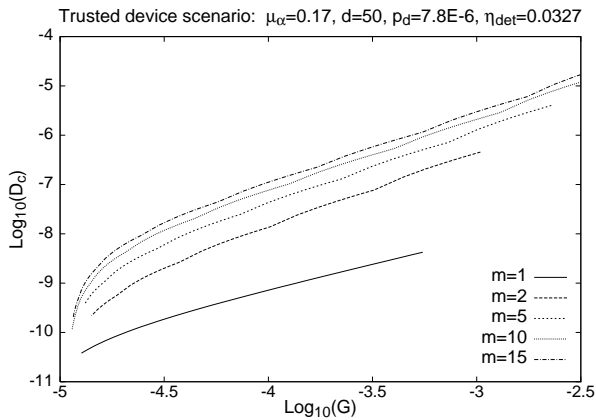}}}
\end{center}
\caption{Gain ($G$) versus double click rate ($D_c$) in a sequential attack for
different values of the photon number $m$, and for the 
optimal distribution of the state coefficients $A_{n,m}^{(k)}$
derived in Sec.~\ref{susda}. The experimental parameters 
coincide with those used in Fig.~\ref{exp3tb}. \label{exp3Dtb}}
\end{figure}

As already suggested in Ref.~\cite{curty_dps}, in this QKD protocol
it is not enough for Alice and Bob to include the effect of the double clicks 
obtained by Bob in the QBER, but it might be very useful for the legitimate 
users to monitor also the double click rate to guarantee security against a 
sequential attack. Fig.~\ref{exp3Dtb} shows a graphical representation 
of the gain versus the double click rate for the case where 
$\mu_{\alpha}=0.17$, $\tilde{d}=50$, $p_d=7.8\times{}10^{-6}$, $\eta_{det}=0.0327$, 
and for different values of the parameter $m$ \cite{dpsqkd_exp1}. 
(See also Ref.~\cite{dpsqkd_exp2b}.) Similar results can also be obtained 
for the experimental parameters used in Refs.~\cite{dpsqkd_exp2,dpsqkd_exp3}.
As expected, the double click rate at Bob's side decreases as the losses 
in the channel increases and the photon number $m$ decreases. If Alice 
and Bob only accept a double click rate below the curve which corresponds 
to the case $m=1$ then they could always detect the sequential attacks presented 
in this section.

\section{CONCLUSION}\label{CONC}

In this paper we have quantitatively analyzed the effect that Bob's detectors dead-time
has on the performance of sequential attacks
against a differential-phase-shift (DPS) quantum key distribution (QKD) protocol
based on weak coherent pulses. A sequential attack consists of Eve measuring out 
every coherent state emitted by Alice and, afterwards, she prepares new signal 
states, depending on the results obtained, that are given to Bob. Whenever 
Eve obtains a predetermined number of consecutive successful measurement outcomes, 
then she prepares a new train of non-vacuum signal states that is forwarded to Bob. Otherwise,
Eve can send vacuum signals to Bob to avoid errors. Sequential attacks transform 
the original quantum channel between Alice and Bob into an entanglement breaking 
channel and, therefore, they do not allow the distribution of quantum correlations 
needed to establish a secret key. 

Specifically, we have studied sequential attacks where Eve realizes unambiguous
state discrimination of Alice's signal states. When Eve identifies 
unambiguously a signal state sent by Alice, then she considers this result 
as successful. Otherwise, she considers it as a failure. Moreover, we have 
considered two possible scenarios for our analysis. In the
first one, so-called untrusted device scenario, we assumed that
Eve can control some imperfections in Alice and Bob's devices ({\it
e.g.}, the detection efficiency, the dark count probability, 
and the dead-time of Bob's detectors), together with 
the quantum channel, and she exploits them to obtain maximal
information about the shared key. In the second scenario,
so-called trusted device scenario, we considered that Eve
cannot modify the actual detection devices employed by Alice and
Bob. That is, the legitimate users have complete knowledge
about their detectors, which are fixed by the actual experiment.
From a practical point of view, this last case
constitutes a reasonable description of a realistic situation,
where Alice and Bob could in principle try to limit Eve's influence on their apparatus
by some counterattack techniques.

As a result, we obtained upper bounds on the maximal distance achievable 
by a DPS QKD scheme as a function of the error rate in the sifted key, and 
the detection efficiency, 
the dark count probability, and the dead-time of Bob's detectors. It states 
that no key distillation protocol can provide a secret key from the 
correlations established by the users. While our analysis seems to indicate 
that in the untrusted device scenario
all the long-distance implementations of DPS QKD reported so far in 
the literature would be insecure against a sequential attack, 
it also suggests that, in the trusted device scenario, 
it might be very useful for the legitimate users to monitor also the 
double click rate at Bob's side. This fact might dramatically increase Alice and Bob's
ability in defeating sequential attacks in this case.  

\section{ACKNOWLEDGEMENTS}

The authors wish to thank Norbert L\"utkenhaus, Bing Qi and especially 
Hoi-Kwong Lo for very fruitful discussions on the topic of this 
paper and very useful comments on the manuscript. Financial support 
from DFG under the Emmy Noether programme, and the European Commission 
(Integrated Project SECOQC) are gratefully acknowledged. This research was 
supported in part by Perimeter Institute for Theoretical Physics. Research 
at Perimeter Institute is supported in part by the Government of Canada 
through NSERC and by the province of Ontario through MEDT. M.C. and K.T.
would like to thank Hoi-Kwong Lo for hospitality and support during their 
stays at the University of Toronto. K.T. also thanks the National
Institute of Information and Communications Technology, in Japan, for 
the support.
  
\appendix

\section{Average total number of errors $e(k)$}\label{ap_a}

In this appendix we obtain an expression for the average total number of 
errors $e(k)$ obtained by Bob when Eve
sends him a signal state $\ket{\psi_e^k}$ followed by $1+d$ vacuum
states (Cases A, B, and C in Fig.~\ref{strategy1_meas}). 
These signal states always produce one single click in Bob's 
detection apparatus and, therefore, they can cause at most one 
single error ({\it i.e.}, $e(k)\leq{}1$). 
This error can occur in any temporal mode $n\in[0,k]$. The parameter 
$e(k)$ can be written as
\begin{equation}\label{una}
e(k)=\sum_{n=0}^{k} p_{err}^{k,n},
\end{equation}
where $p_{err}^{k,n}$ denotes the probability that Bob 
obtains an error in temporal mode $n$. Next, we calculate this 
probability. 

We shall consider that Bob employs the detection setup shown in
Fig.~\ref{dpsqkd}. We will assume as well that his detectors, $D0$ and
$D1$, have a detection efficiency equal to one, a dark count probability 
equal to zero, and they cannot distinguish the
number of photons of arrival signals. That is, they provide only two
possible outcomes: ``click" (at least one photon is detected), and
``no click" (no photon is detected in the pulse). The action of $D0$ and
$D1$ in the time slot $n$ can be characterized by one
positive operator value measure (POVM) that contains four
elements: $D_{vac}^n$, $D_0^n$, $D_1^n$, and $D_{Dc}^n$. The outcome of
the first operator $D_{vac}^n$ corresponds to no click in the
detectors, the following POVM operator $D_0^n$ ($D_1^n$) gives
precisely one detection click in detector $D0$ ($D1$), and the 
last one $D_{Dc}^n$ gives rise to both
detectors being triggered. If we denote by $\ket{p,q}_{D0,D1}^n$ the
state that, in temporal mode $n$, contains $p$ photons in the spatial mode arriving to
detector $D0$ and $q$ photons in the spatial mode corresponding
to detector $D1$, then the elements of this POVM can be expressed as
\begin{eqnarray}\label{povm_detectors1}
D_{vac}^n=\ket{0,0}\bra{0,0}^n_{D0,D1}\nonumber\\
D_0^n=\sum_{p=1}^{\infty}\ket{p,0}\bra{p,0}^n_{D0,D1}\nonumber\\
D_1^n=\sum_{p=1}^{\infty}\ket{0,p}\bra{0,p}^n_{D0,D1}\nonumber\\
D_{Dc}^n=\sum_{p,q=1}^{\infty}\ket{p,q}\bra{p,q}^n_{D0,D1}
\end{eqnarray}

Once the state 
$\ket{\psi_e^k}$ followed by $1+d$ vacuum
states passes Bob's interferometer, the 
signal that arrives at Bob's detectors, 
that we shall denote as $\ket{\tilde{\psi}_e^k}$, is given by 
\begin{equation}\label{arriving_state1}
\ket{\tilde{\psi}_e^k}=\sum_{n=1}^k
B_n^{(k)} \big[\hat{a}_{n,D0}^\dag-\hat{a}_{n,D1}^\dag
+\hat{a}_{n-1,D0}^\dag+\hat{a}_{n-1,D1}^\dag\big]
\ket{0}
\end{equation}
followed by $d$ vacuum states. The coefficients $B_n^{(k)}$ 
in Eq.~(\ref{arriving_state1}) are 
given by $B_n^{(k)}=[A_n^{(k)} \exp{(i\theta_n)}]/2$,
and $\hat{a}_{n,D0}^\dag$ ($\hat{a}_{n,D1}^\dag$) represents a
creation operator for one photon in temporal mode $n$ and in the
spatial mode corresponding to detector $D0$ ($D1$). 

It turns out that the probability $p_{err}^{k,n}$ remains constant independently of the
different possible combinations of correct phases $\theta_n$ identified by 
Eve. Therefore, without loss of generality, we can assume a fixed 
value for the angles $\theta_n$. In particular, 
we shall consider, for instance, that $\theta_n=0$ for all $n\in[1,k]$ and, 
consequently,   
$B_n^{(k)}=A_n^{(k)}/2$. In this scenario an error occurs when detector $D1$ clicks. 
The probability $p_{err}^{k,n}$ can then be expressed as
$p_{err}^{k,n}=\text{Tr}(D_1^n \ket{\tilde{\psi}_e^k}\bra{\tilde{\psi}_e^k})$, 
with $\ket{\tilde{\psi}_e^k}$ given by Eq.~(\ref{arriving_state1}).
As a result, we obtain 
\begin{eqnarray}\label{p_error_last}
p_{err}^{k,k}&=&\frac{1}{4}\vert{}A_k^{(k)}\vert{}^2\nonumber\\
p_{err}^{k,0<n<k}&=&\frac{1}{4}\vert{}A_{n+1}^{(k)}-A_n^{(k)}\vert{}^2\nonumber\\
p_{err}^{k,0}&=&\frac{1}{4}\vert{}A_1^{(k)}\vert{}^2.
\end{eqnarray}

Adding all these terms together according to Eq.~(\ref{una})
we finally obtain
\begin{equation}\label{eq_ek_ap}
e(k)=\frac{1}{4}\Bigg[\vert{}A_1^{(k)}\vert{}^2+\sum_{n=1}^{k-1}
\vert{}A_{n+1}^{(k)}-A_n^{(k)}\vert{}^2+\vert{}A_k^{(k)}\vert{}^2\Bigg].
\end{equation}

\section{Optimization of $A_n^{(k)}$}\label{ap_opt}

In this appendix we describe a method to optimize the state coefficients $A_n^{(k)}$
introduced in Eq.~(\ref{signal_uds}) for  
the untrusted device scenario. As shown in Sec.~\ref{gain_A}, in this case the overall gain $G$
is independent of the chosen distribution for these 
coefficients. This means that the minimum value of the 
QBER can be attained by optimizing 
the average total number of errors $e(k)$ independently of $G$. 

The coefficients $A_n^{(k)} \in \mathbb{C}$ satisfy the
normalization condition $\sum_{n=1}^k
\vert{}A_n^{(k)}\vert{}^2=1$. 
In polar coordinates each of these coefficients can 
be expressed as $A_n^{(k)}=a^{(k)}_n \exp(i\psi^{(k)}_n)$, 
with $a^{(k)}_n$, and $\psi^{(k)}_n \in \mathbb{R}$. With 
this notation, the terms $|A_{n+1}^{(k)}-A_{n}^{(k)}|^2$ 
in Eq.~(\ref{eq_ek}) simplify to
\begin{eqnarray}
|A_{n+1}^{(k)}-A_{n}^{(k)}|^2&=&(a^{(k)}_{n+1})^2+(a^{(k)}_{n})^2 \nonumber \\
&-&2a^{(k)}_{n} a^{(k)}_{n+1} \cos(\psi^{(k)}_{n+1}-\psi^{(k)}_n) \nonumber \\
&\geq& (a^{(k)}_{n+1}-a^{(k)}_{n})^2,
\end{eqnarray}
where equality is obtained iff $\psi^{(k)}_{n+1}-\psi^{(k)}_n$ is an even multiple 
of $\pi$, {\it i.e.}, both $A_{n+1}^{(k)}$ and $A_{n}^{(k)}$ have the same phase. Then, without loss 
of generality, we can always impose the phases 
$\psi^{(k)}_n$, with $n\in[1,k]$, to be equal to zero. With this 
constraint, together with 
the normalization condition of the state coefficients, one can further simplify the 
parameters $e(k)$ as
\begin{equation}
\label{eq: e(k1)}
e(k)=\frac{1}{2} \left(1-\sum_{n=1}^{k-1} a_{n+1}^{(k)} a_{n}^{(k)} \right).
\end{equation}
Now, in order to minimize this quantity one can equivalently solve the following 
optimization problem,
\begin{eqnarray}
\label{eq: opt1}
\text{minimize} &&c \sum_{n=1}^k (a_{n}^{(k)})^2 - \sum_{n=1}^{k-1} a_{n+1}^{(k)} a_{n}^{(k)} \nonumber \\
\text{subject to} && \sum_{n=1}^{k} (a_{n}^{(k)})^2=1,
\end{eqnarray}
where $c>0$ is an arbitrary positive constant. 
This optimization problem can be written in matrix form as
\begin{equation}
\text{minimize}_{\| \vec a_k \|=1} f(\vec a_k) = \vec a_k M(c) \vec a_k^T, 
\end{equation}
where the vector $\vec a_k$ is defined as $\vec a_k=( a_1^{(k)}, \dots,  a_k^{(k)})$, 
and the matrix $M(c)=c\openone-K/2$ with $K=K^T$ being a $k \times k$ matrix 
with ones only on the first off-diagonals and zeros elsewhere, {\it i.e.},
\begin{equation}
K =
\left( \begin{array}{ccccccc}
0 & 1 & 0 & \ldots \\
1 & 0 & 1 & \ldots \\
0 & 1 & 0 & \ldots \\
\vdots & \vdots & \vdots & \ddots 
\end{array} \right).
\end{equation}
The Hessian matrix of the function $f(\vec a_k)$ is $M(c)$; hence if one selects the constant 
$c$ such that $M(c) > 0$ then the resulting optimization problem is convex, and it suffices to 
find a local minimum. The solution is given by
\begin{eqnarray}
\text{minimize}_{\| \vec a_k \|=1}f(\vec a_k)&=&c-\text{maximize}_{\| 
\vec a_k \|=1} \vec a_k K \vec a_k \nonumber \\
&=& c - \lambda_{max}(K),
\end{eqnarray}
where $\lambda_{max}(K)$ denotes the maximal eigenvalue of the matrix $K$. The 
optimal solution for the state coefficients $A_n^{(k)}$ coincides then 
with the elements of the normalized eigenvector of $K$ that is 
associated with its maximal eigenvalue.

\begin{figure}
\centering
\includegraphics[scale=0.305,angle=-90]{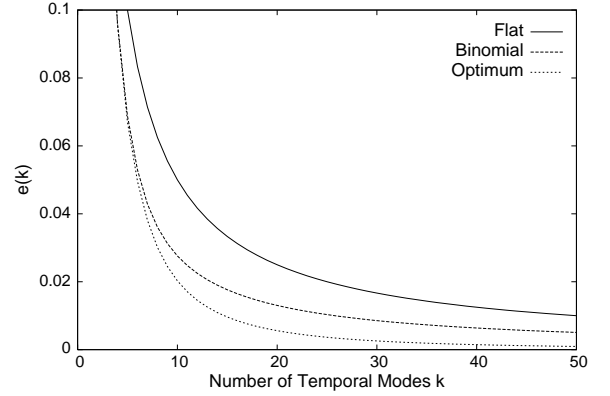} 
\caption{The average total number of errors $e(k)$
versus $k$ for
different distributions of the state coefficients $A_n^{(k)}$: 
flat (solid), binomial (dashed), and the optimal distribution
(dotted).}
\label{fig: e(k)}
\end{figure}

Fig.~\ref{fig: e(k)} shows a 
graphical representation of $e(k)$
versus $k$ for the optimal distribution of $A_n^{(k)}$ together with the flat and the 
binomial distributions given in Eq.~(\ref{distribu}). 

\section{Probabilities $q(k)$, $r(k)$ and $s(k)$}\label{ap_b}

In this appendix we provide the expressions for the
a priori probabilities of the blocks of signal states 
illustrated in Fig.~\ref{strategy1_meas_dead_time}: 
$q(k)$, $r(k)$ and $s(k)$. As already explained in Sec.~\ref{medattack}, 
these blocks of signal states arise due to the effect of the dead-time 
of Bob's detectors. In particular, in the trusted device scenario
it is not guaranteed that the first temporal mode that arrives 
at Bob's side once his detectors are recovered from a dead-time 
coincides with the first temporal mode of any of the blocks of signals 
considered in Fig.~\ref{strategy1_meas}. In this scenario, and for a 
given block 
of signals containing $k+1$ modes (see Fig.~\ref{strategy1_meas}),
we shall denote with  
$p_d(n)$ the probability 
that the first temporal mode of the block that arrives at Bob's side after a dead-time 
is mode $k-n$.
That is, $p_d(n)$ 
represents the probability that a dead-time finishes 
once Bob has already received the first $n$ temporal modes of a given
block of signals. 
For instance, $p_d(0)$ represents the probability that the first 
mode that arrives at Bob's side after a dead-time coincides with the first temporal 
mode of the block, $p_d(1)$ denotes 
the probability that the first mode arriving at Bob's side corresponds to the 
second temporal mode of the block,
and so on. Here
we use again the labeling convention illustrated in Fig.~\ref{ordering}.

With this notation, let us now calculate the parameter $q(0)$, {\it i.e.}, the 
probability that Bob receives a vacuum state after a dead-time 
(Case A in Fig.~\ref{strategy1_meas_dead_time}). In principle,
this vacuum 
state could originate from the last temporal mode of every block 
of signal states illustrated in Fig.~\ref{strategy1_meas} with $d=0$. 
For instance, the probability that it originates from the block of signals 
which contains only one vacuum state 
(Case D in Fig.~\ref{strategy1_meas} when the parameter $k=0$) is given by
$p_d(0)p_v(0)$, where $p_v(0)$ is given by Eq.~(\ref{pv_k}). 
In general, we have that the probability that this vacuum 
state arises from a block of signals which contains $k+1$ vacuum states, 
with $0\leq{}k\leq{}M_{min}$
(Cases D and E in Fig.~\ref{strategy1_meas}), is given by
$p_d(k)p_v(k)/[\sum_{m=n-k}^{M_{min}}p_v(m)+\sum_{m'=M_{min}}^{M_{max}}p_s(m')]$,
where 
the normalization factor $\sum_{m=n-k}^{M_{min}}p_v(m)
+\sum_{m'=M_{min}}^{M_{max}}p_s(m')$ is used to guarantee that, 
for each value of the parameter $p_d(n)$, the probabilities of the different blocks of
signal states which contain at least $n+1$ temporal modes add one.    
Similarly, we find that the probability that this vacuum state originates 
from a block of signals containing a state $\rho_{e}^k$ followed 
by one vacuum state, with $M_{min}\leq{}k\leq{}M_{max}$
(Cases A, B, and C in Fig.~\ref{strategy1_meas} with $d=0$),
can be written as $p_d(k)p_s(k)/\{\sum_{m=k}^{M_{max}}[p_v(m)+p_s(m)]\}$. 
After adding all these terms together, we obtain that 
$q(0)$ can be expressed as
\begin{eqnarray}\label{q0}
q(0)&=&\sum_{n=0}^{M_{min}}\frac{p_d(n)p_v(n)}
{1-\sum_{m=0}^{n-1}p_v(m)} \nonumber \\
\ &+&\sum_{k=M_{min}}^{M_{max}}\frac{p_d(k)p_s(k)}{\sum_{m=k}^{M_{max}}\big[p_v(m)+p_s(m)\big]} 
\end{eqnarray}

The analysis of the remaining cases is completely analogous. In particular, we find that
\begin{equation}
q(k)=\sum_{n=k}^{M_{min}} \frac{p_d(n-k)p_v(n)}{1-\sum_{m=0}^{n-1-k}
p_v(m)},
\end{equation}
with $1\leq{}k\leq{}M_{min}$,
\begin{equation}\label{rk}
r(k)=\sum_{n=M_{min}}^{M_{max}}
\frac{p_d(n-1-k)p_s(n)}{\sum_{m=n-1-k}^{M_{max}}\big[p_v(m)+p_s(m)\big]},
\end{equation}
with $0\leq{}k\leq{}M_{min}-2$, and 
\begin{equation}\label{sk}
s(k)=\sum_{n=M_{min}+k}^{M_{max}} \frac{p_d(n-M_{min}-k)
p_s(n)}{\sum_{m=n-M_{min}-k}^{M_{max}}\big[p_v(m)+p_s(m)\big]},
\end{equation}
with $0\leq{}k\leq{}M_{max}-M_{min}$.

To conclude, let us present very briefly a numerical method to calculate the probabilities $p_d(n)$, 
with $0\leq{}n\leq{}M_{max}$. 
This procedure is used in Sec.~\ref{evalua2} for the graphical representation
of the gain $G$ versus the QBER (and also of the double click rate versus $G$) 
in a sequential attack. For simplicity, let us consider the 
situation where $M_{max}=\lceil{}t_df_c\rceil$.
That is, each block of signal states illustrated in Fig.~\ref{strategy1_meas_dead_time}
could produce, at most, only one single click in Bob's detectors due to 
the effect of the dead-time.  
In this scenario, the probabilities $p_d(n)$ only depend on whether 
the previous $M_{max}+1$ temporal modes which precede a block of signals 
clicked or did not click. Especifically, 
we have that $p_d(0)$ is given by the probability that none of these $M_{max}+1$ 
previous modes clicked or it only clicked the first of them. That is, 
\begin{equation}\label{casi1}
p_d(0)=G+(1-G)^{M_{max}+1}.
\end{equation}
Similarly, $p_d(n)$, with $1\leq{}n\leq{}M_{max}$, can be expressed as
\begin{equation}\label{casi2}
p_d(n)=G(1-G)^{n}.
\end{equation}
Then, in order to obtain $p_d(n)$ an plot the figures included in Sec.~\ref{evalua2}, 
we perform several iterations 
for each value of the gain $G$. First,
we calculate the probabilities $p_d(n)$ according to Eq.~(\ref{casi1}) and 
Eq.~(\ref{casi2}) and, with these probabilities, we obtain a new value for the gain
$G$. 
We repeat this procedure several times until the value of $G$ converges to a constant
value.

\section{Probabilities $p_{vv}$, $p_{vk}^m$ and $p_{c,\bar{k},k}^m(n)$}\label{ap_c}

In this appendix we obtain an expression for the probabilities of
obtaining a click at Bob's side: $p_{vv}$, $p_{vk}^m$ and $p_{c,\bar{k},k}^m(n)$, 
introduced in Sec.~\ref{gain_A_tds}.

Let us start by describing Bob's detectors $D0$ and $D1$ in the trusted
device scenario. As already introduced in Sec.~\ref{medattack}, 
these detectors are characterized by their detection efficiency
$\eta_{det}$, their dark count probability $p_d$, and their dead-time. 
The detection efficiency of $D0$ and $D1$ can be modeled by a combination of beam
splitters of transmittance $\eta_{det}$ and {\it ideal} detectors
\cite{yurke_85}. This model can be simplified further by
considering that both detectors have the same detection efficiency. In this situation, it
is possible to attribute the losses of both detectors to a
single-loss beam splitter which is located after the transmission
channel. Moreover, like in Sec.~\ref{susda}, we shall assume that $D0$ 
and $D1$ cannot distinguish the
number of photons of arrival signals, but they provide only as 
possible outcomes ``click" and
``no click". This means, in particular, that the action of Bob's ideal detectors can be 
characterized by the POVM given by Eq.~(\ref{povm_detectors1}). 
Furthermore, the noise introduced by  
$D0$ and $D1$ due to their dark counts can be considered to be
independent of the incoming signals.
Note that the clicks that Bob observes can be thought as coming from a
two-step process: in the first step the signal states sent by Eve 
can produce clicks in his detectors, while in the
second step random clicks from the detector's dark counts is added.

Next, we calculate an expression for the probability $p_{c,\bar{k},k}^m(n)$. For that, we 
obtain first the probability that Bob's detectors fire 
due to the signal states sent by Eve only, {\it i.e.}, without taking into
account the dark counts of $D0$ and $D1$; afterwards we will 
include in the analysis the effect of the dark counts of the detectors. 

Once a state 
$\ket{\psi_k^m}$, followed by one vacuum
state, passes Bob's interferometer together with the beam splitter 
which models the losses of his detectors, the 
signal that arrives at Bob's ideal detectors, 
that we shall denote as $\ket{\tilde{\psi}_k^m}$, is given by 
\begin{equation}\label{arriving_state}
\ket{\tilde{\psi}_k^m}=\frac{1}{\sqrt{m!}} \Big[\sum_{n=0}^k
(E_n^m \hat{a}_{n,D1}^\dag+F_n^m \hat{a}_{n,D0}^\dag)+\sum_{n=1}^k
G_n^m \hat{b}_{n}^\dag\Big]^m\ket{0},
\end{equation}
where $\hat{a}_{n,D0}^\dag$ ($\hat{a}_{n,D1}^\dag$) represents 
again a creation operator for one photon in temporal mode $n$ and in the
spatial mode arriving to detector $D0$ ($D1$), $\hat{b}_{n}^\dag$
denotes a creation operator for one photon in temporal mode $n$ and in the
spatial mode that goes out of the beam splitter that models the losses
of the detectors, and the
coefficients $E_n^m$, $F_n^m$, and $G_n^m$ are given by
\begin{equation}\label{ees}
E_n^m = \left\{ \begin{array}{ll} \eta
A_{1,m}^{(k)} e^{i\theta_1} & \textrm{if $n=0$}\\
\eta[A_{n+1,m}^{(k)} e^{i\theta_{n+1}}-A_{n,m}^{(k)} e^{i\theta_n}]
& \textrm{if $1\leq{}n\leq{}k-1$}\\
-\eta A_{k,m}^{(k)} e^{i\theta_k} & \textrm{if $n=k$},\\
\end{array} \right.
\end{equation}
with $\eta=\sqrt{\eta_{det}}/2$, 
\begin{equation}\label{efes}
F_n^m = \left\{ \begin{array}{ll} \eta
A_{1,m}^{(k)} e^{i\theta_1} & \textrm{if $n=0$}\\
\eta[A_{n+1,m}^{(k)} e^{i\theta_{n+1}}+A_{n,m}^{(k)} e^{i\theta_n}]
& \textrm{if $1\leq{}n\leq{}k-1$}\\
\eta A_{k,m}^{(k)} e^{i\theta_k} & \textrm{if $n=k$},\\
\end{array} \right.
\end{equation}
and
\begin{equation}\label{gees}
G_n^m=\sqrt{1-\eta_{det}} A_{n,m}^{(k)} e^{i\theta_n},
\end{equation}
respectively.

Let $p_{c,\bar{k},k}^{m,s}(n)$ denote the probability that Bob 
obtains a click in mode $n$, with $0\leq{}n\leq{}\bar{k}$, and he
does not obtain a click in any previous temporal mode $l$, with $n<l\leq{}\bar{k}$, 
when he receives 
the last $\bar{k}$
temporal modes of a signal state $\ket{\psi_{k}^m}$ followed by one 
vacuum state. This probability can be expressed as
\begin{equation}\label{clicks_signals}
p_{c,\bar{k},k}^{m,s}(n)=p_{\bar{k},k,D0}^{m}(n)+p_{\bar{k},k,D1}^{m}(n)
+p_{\bar{k},k,D_{Dc}}^{m}(n),
\end{equation}
where $p_{\bar{k},k,D0}^{m}(n)$ ($p_{\bar{k},k,D1}^{m}(n)$) represents
the probability that only detector $D0$ ($D1$) clicks in temporal
mode $n$, and $p_{\bar{k},k,D_{Dc}}^{m}(n)$ denotes the probability that
both detectors $D0$ and $D1$ click in temporal mode $n$ \cite{nuevotexto}.  
These probabilities are given by 
\begin{equation}\label{auxiliar}
p_{\bar{k},k,Di}^{m}(n)=\text{Tr}\bigg(\bigotimes_{l=n+1}^{\bar{k}} D_{vac}^l\otimes{}D_i^n
\ket{\tilde{\psi}_k^m}\bra{\tilde{\psi}_k^m}\bigg), 
\end{equation}
where the POVM elements $D_{vac}^l$ and 
$D_i^n$, with $i\in{}\{0,1,D_c\}$, are given by Eq.~(\ref{povm_detectors1}).
After some calculations, we obtain that these probabilities 
can be written, respectively, as

\begin{eqnarray}\label{pc0}
p_{\bar{k},k,D0}^{m}(n)&=&\sum_{\substack{\sum_{\substack{r=0\\
r\notin{}I_0}}^{3k+1}t_r=m
\\ t_{(k+n+1)}\geq{}1}}
\frac{m!} {\prod_{\substack{l=0\\ l\notin{}I_0}}^{3k+1}t_l!}
\prod_{\substack{i=0\\ i\notin{}[n,\bar{k}]}}^{k}
\vert{}E_i^m\vert{}^{2t_i}\nonumber \\
\ &\times&\prod_{\substack{j=0\\ j\notin{}[n+1,\bar{k}]}}^{k}
\vert{}F_j^m\vert{}^{2t_{(j+k+1)}}
\prod_{s=1}^{k}
\vert{}G_s^m\vert{}^{2t_{(s+2k+1)}}, \nonumber \\
p_{\bar{k},k,D1}^{m}(n)&=&\sum_{\substack{\sum_{\substack{r=0\\r\notin{}I_1}}^{3k+1}t_r=m
\\ t_{n}\geq{}1}}
\frac{m!} {\prod_{\substack{l=0\\ l\notin{}I_1}}^{3k+1}t_l!}
\prod_{\substack{i=0\\ i\notin{}[n+1,\bar{k}]}}^{k} \vert{}E_i^m\vert{}^{2t_i}\nonumber \\
\ &\times&\prod_{\substack{j=0\\ j\notin{}[n,\bar{k}]}}^{k}
\vert{}F_j^m\vert{}^{2t_{(j+k+1)}}
\prod_{s=1}^{k}
\vert{}G_s^m\vert{}^{2t_{(s+2k+1)}}, \nonumber \\
p_{\bar{k},k,D_{Dc}}^{m}(n)&=&\sum_{\substack{\sum_{\substack{r=0\\r\notin{}I_{Dc}}}^{3k+1}t_r=m
\\ t_{(k+n+1)}\geq{}1 \\ t_{n}\geq{}1}}
\frac{m!} {\prod_{\substack{l=0\\ l\notin{}I_{Dc}}}^{3k+1}t_l!}
\prod_{\substack{i=0\\ i\notin{}[n+1,\bar{k}]}}^{k} \big[\vert{}E_i^m\vert{}^{t_i} \nonumber \\
\ &\times&\vert{}F_i^m\vert{}^{t_{(i+k+1)}}\big]^2
\prod_{s=1}^{k}
\vert{}G_s^m\vert{}^{2t_{(s+2k+1)}},
\end{eqnarray}
where the sets of indexes $I_0$, $I_1$ and $I_{Dc}$ are given by: 
$I_0=[n,\bar{k}]\cup[n+k+2,\bar{k}+k+1]$,
$I_1=[n+1,\bar{k}]\cup[n+k+1,\bar{k}+k+1]$ and
$I_{Dc}=[n+1,\bar{k}]\cup[n+k+2,\bar{k}+k+1]$.
It turns out that the probability $p_{c,\bar{k},k}^{m,s}(n)$ remains invariable 
independently of the
different possible combinations of correct consecutive phases $\theta_n$ identified by 
Eve. Therefore, without loss of generality, we can always assume a fixed 
value for the angles $\theta_n$; for instance, we can consider that 
$\theta_n=0$ for all $n\in[1,k]$. This means, in particular, that the parameters
$E_n^m$, $F_n^m$ and $G_n^m$ that appear in Eq.~(\ref{pc0}) depend only 
on the state coefficients $A_{n,m}^{(k)}$ and the detection efficiency of 
Bob's detectors.  

In order to include the effect of the dark counts of $D0$ and $D1$ 
in the analysis, let us define the parameter $P_d$ as
\begin{equation}\label{par_Pd}
P_d=p_d(2-p_d).
\end{equation}
This is the probability to have a click at Bob's side in a given time 
slot due to the 
dark counts of his detectors only. As already mentioned in Sec.~\ref{gain_A_tds}, 
here we consider that double click events are not discarded by
Bob. Every time Bob obtains a
double click, he just decides randomly the bit value \cite{Norbert99}.
With this notation, it turns out that the probability that Bob 
obtains a click in temporal mode $n$, with $0\leq{}n\leq{}\bar{k}$, and he
does not obtain a click in any previous mode $l$, with $n<l\leq{}\bar{k}$, 
due to the dark counts of his detectors only, probability that we shall denote as 
$p_{c,\bar{k}}^{m,d}(n)$, is given by
\begin{equation}\label{clicks_dark}
p_{c,\bar{k}}^{m,d}(n)=P_d(1-P_d)^{\bar{k}-n}.
\end{equation}

Combining Eq.~(\ref{clicks_signals}) and Eq.~(\ref{clicks_dark}), we obtain that the 
probability $p_{c,\bar{k},k}^m(n)$ can be expressed as
\begin{eqnarray}\label{psv}
p_{c,\bar{k},k}^m(n)&=&p_{c,\bar{k},k}^{m,s}(n)(1-P_d)^{\bar{k}-n+1}+p_{c,\bar{k}}^{m,d}(n) \\
\ &\times&p_{\bar{k},k,D_{vac}}^{m}(n)+p_{c,\bar{k},k}^{m,s}(n)p_{c,\bar{k}}^{m,d}(n) \nonumber \\
\ &=&p_{c,\bar{k},k}^{m,s}(n)(1-P_d)^{\bar{k}-n}+p_{c,\bar{k}}^{m,d}(n)p_{\bar{k},k,D_{vac}}^{m}(n), \nonumber 
\end{eqnarray}
where $p_{\bar{k},k,D_{vac}}^{m}(n)$ is given by Eq.~(\ref{auxiliar}) with
$D_i=D_{vac}$, {\it i.e.}, it represents the probability that 
Bob does not obtain a click in any temporal mode $l$, with $n\leq{}l\leq{}\bar{k}$, 
due to the signals sent by Eve only. 
This probability can be written as
\begin{eqnarray}\label{pcvac}
p_{\bar{k},k,D_{vac}}^{m}(n)&=&\sum_{\substack{\sum_{\substack{r=0\\r\notin{}I_{vac}}}^{3k+1}t_r=m}}
\frac{m!} {\prod_{\substack{l=0\\ l\notin{}I_{vac}}}^{3k+1}t_l!}
\prod_{\substack{i=0\\ i\notin{}[n,\bar{k}]}}^{k} \big[\vert{}E_i^m\vert{}^{t_i} \nonumber \\
\ &\times&\vert{}F_i^m\vert{}^{t_{(i+k+1)}}\big]^2
\prod_{s=1}^{k}
\vert{}G_s^m\vert{}^{2t_{(s+2k+1)}},
\end{eqnarray}
with the set $I_{vac}=[n,\bar{k}]\cup[n+k+1,\bar{k}+k+1]$. 
 
Finally, we obtain an expression for the probabilities $p_{vv}$ and $p_{vk}^m$. 
The first one, $p_{vv}$, represents the
probability that Bob obtains a click when he receives a vacuum
state and the preceding signal is also a vacuum state. This
probability has the form $p_{vv}=P_d$, 
where $P_d$ is given by Eq.~(\ref{par_Pd}). Similarly, $p_{vk}^m$ 
denotes the probability that Bob obtains a click when he receives a vacuum state and the
preceding signal is the state $\ket{\psi_k^m}$. This probability
is given by $p_{vk}^m=p_{c,0,k}^m(0)$, where $p_{c,\bar{k},k}^m(n)$ is defined in
Eq.~(\ref{psv}). 

\section{Probabilities $p_{pv}$ and $p_{pk}$}\label{ap_ppv}

In this appendix we obtain an expression for the probabilities $p_{pv}$ and $p_{pk}$. 
Let us begin with $p_{pv}$, {\it i.e.}, the probability that the 
signal which precedes a vacuum state (Case A in Fig.~\ref{strategy1_meas_dead_time}) 
is also a vacuum state.

As shown in Appendix~\ref{ap_b}, after a dead-time Bob receives a vacuum state 
with a probability given by Eq.~(\ref{q0}). The first summation in Eq.~(\ref{q0})
refers to the probability that this vacuum state originates from a block of 
signal states containing only vacuum pulses (Cases D and E 
in Fig.~\ref{strategy1_meas}). We have, therefore, that $p_{pv}$
can be written as
\begin{equation}
p_{pv}=\frac{1}{q(0)}\sum_{n=0}^{M_{min}}\frac{p_d(n)p_v(n)}
{1-\sum_{m=0}^{n-1}p_v(m)},
\end{equation} 
where the normalization factor $q(0)$ is used to guarantee that 
$p_{pv}+\sum_{M_{min}}^{M_{max}}p_{pk}=1$. Similarly, each term $p_d(k)p_s(k)/\sum_{m=k}^{M_{max}}\big[p_v(m)+p_s(m)\big]$ that 
appears in the second summation of Eq.~(\ref{q0}) 
represents the probability that Bob receives a vacuum state preceded by a 
signal $\ket{\psi_{k}^m}$ (Cases A, B and C in Fig.~\ref{strategy1_meas}
when the number of photons contained in $\rho_e^k$ is equal to $m$). After normalizing 
by the factor $q(0)$ we find that $p_{pk}$ is given by
\begin{equation}
p_{pk}=\frac{p_d(k)p_s(k)}{q(0)\sum_{m=k}^{M_{max}}\big[p_v(m)+p_s(m)\big]},
\end{equation} 
with $M_{min}\leq{}k\leq{}M_{max}$.

\section{Probability $p^k_{\bar{k}}$}\label{ap_pkkb}

In this appendix we calculate the probability $p^k_{\bar{k}}$ with 
$1\leq{}\bar{k}\leq{}M_{min}-1$, 
{\it i.e.}, the probability that $\rho_e^{\bar{k}}$ represents the last $\bar{k}$
temporal modes of a signal state $\ket{\psi_{k}^m}$ with 
$M_{min}\leq{}k\leq{}M_{max}$ (Cases
B, D, and F in Fig.~\ref{strategy1_meas_dead_time}).

As shown in Appendix~\ref{ap_b}, after a dead-time Bob receives the 
signal $\rho_e^{\bar{k}}$ followed by one vacuum state with probability 
$r(\bar{k}-1)$, where the probability $r(k)$ is 
given by Eq.~(\ref{rk}). Each term 
$p_d(n-1-l)p_s(n)/\sum_{m=n-1-l}^{M_{max}}[p_v(m)+p_s(m)]$ that appears 
in the summation of Eq.~(\ref{rk}) denotes the probability that 
Bob receives a state $\rho_e^{l+1}$ which corresponds to the 
last $l+1$
temporal modes of a signal $\ket{\psi_{n}^m}$. After substituting $l+1=\bar{k}$ and 
$n=k$, and normalizing 
by the factor $r(\bar{k}-1)$ we find that $p^k_{\bar{k}}$ is given by
\begin{equation}
p_{\bar{k}}^k=\frac{p_d(k-\bar{k})p_s(k)}{r(\bar{k}-1)\sum_{m=k-\bar{k}}^{M_{max}}
[p_v(m)+p_s(m)]}.
\end{equation}

\section{Probabilities $p_{pv\bar{k}}$ and $p_{p\bar{k}}^k$}\label{nac}

In this appendix we obtain an expression for the probabilities $p_{pv\bar{k}}$ 
and $p_{p\bar{k}}^k$, {\it i.e.}, the probability that the signal state 
$\rho_e^{\bar{k}}$ received by Bob, with $M_{min}\leq{}\bar{k}\leq{}M_{max}$,
is preceded by a vacuum state, and the probability that $\rho_e^{\bar{k}}$
represents the last $\bar{k}$ temporal modes of 
$\ket{\psi_{k}^m}$, with
$\bar{k}<k\leq{}M_{max}$, respectively 
(Cases G, H and I in Fig.~\ref{strategy1_meas_dead_time}).

As shown in Appendix~\ref{ap_b}, after a dead-time Bob receives the signal 
state $\rho_e^{\bar{k}}$
with a probability given by Eq.~(\ref{sk}). This probability can be equivalently
written as
\begin{equation}\label{sknueva}
s(\bar{k}-M_{min})=\sum_{n=\bar{k}}^{M_{max}} \frac{p_d(n-\bar{k})
p_s(n)}{\sum_{m=n-\bar{k}}^{M_{max}}\big[p_v(m)+p_s(m)\big]},
\end{equation}  
with $M_{min}\leq{}\bar{k}\leq{}M_{max}$. The first term in the 
summation given by Eq.~(\ref{sknueva}), {\it i.e}, the case 
$n=\bar{k}$, refers to the probability that Bob receives a state 
$\rho_e^{\bar{k}}=\ket{\psi_{\bar{k}}^m}\bra{\psi_{\bar{k}}^m}$. 
Because of the particular structure of 
the blocks of signal states that Eve can send 
to Bob (see Fig.~\ref{strategy1_meas}), these states are always 
preceded by a vacuum state. After normalizing 
by the factor $s(\bar{k}-M_{min})$ we obtain, therefore, that 
$p_{pv\bar{k}}$ is given by
\begin{equation}\label{nse1}
p_{pv\bar{k}}=\frac{p_d(0)p_s(\bar{k})}{s(\bar{k}-M_{min})}.
\end{equation}
Similarly, each term $p_d(n-\bar{k})
p_s(n)/\sum_{m=n-\bar{k}}^{M_{max}}[p_v(m)+p_s(m)]$ that 
appears in Eq.~(\ref{sknueva}), with $\bar{k}<n\leq{}M_{max}$,
represents the probability that Bob receives a state 
$\rho_e^{\bar{k}}$ which originates from the signal state 
$\ket{\psi_{n}^m}$. We find, 
therefore, that 
\begin{equation}\label{nse2}
p_{p\bar{k}}^k=\frac{p_d(k-\bar{k})p_s(k)}{s(\bar{k}-M_{min})\sum_{n=k-\bar{k}}^{M_{max}}
[p_v(n)+p_s(n)]}.
\end{equation}
The normalization factor $s(\bar{k}-M_{min})$ included in 
Eq.~(\ref{nse1}) and in Eq.~(\ref{nse2})
guarantees that 
$p_{pv\bar{k}}+\sum_{k=\bar{k}+1}^{M_{max}} p_{p\bar{k}}^k=1$. 

\section{Probabilities $e_{vv}$,
$e_{vk}^m$ and $p_{e,\bar{k},k}^m(n)$}\label{ap_d}

In this appendix we obtain an expression for the error
probabilities $e_{vv}$, $e_{vk}^m$, and
$p_{e,\bar{k},k}^m(n)$ introduced in Sec.~\ref{qber_a_tds}.

Let us start with the parameter $e_{vv}$, {\it i.e.}, the 
probability that Bob obtains an error
when he receives from Eve a vacuum state and the preceding signal
is also a vacuum state. This quantity is given by 
$e_{vv}=p_d(1-p_d)+p_d^2/2$, where $p_d$ denotes
the dark count probability of Bob's
detectors. This quantity can be further simplified as 
\begin{equation}
e_{vv}=\frac{1}{2}P_d,
\end{equation}
with $P_d$ given by Eq.~(\ref{par_Pd}).  

The parameter $p_{e,\bar{k},k}^m(n)$, {\it i.e.},
the probability that Bob obtains an error in temporal mode $n$ and he 
does not obtain a click in any previous temporal mode $l$, with 
$n<l\leq{}\bar{k}$, when he receives  
the last $\bar{k}$ temporal modes of the signal $\ket{\psi_{k}^m}$
followed by one vacuum state
can be calculated as the probability that Bob obtains a click in the
``wrong" detector and no click in the ``correct" one, together with one
half the probability that he obtains a double click. 
Like in Sec.~\ref{qber_A}, the 
total error probability in this strategy remains invariant independently of the different 
possible combinations of correct phases $\theta_n$ identified by 
Eve. For simplicity, therefore, we can consider again that all these phases 
are equal to zero. This means, in particular, that in this situation 
the ``wrong" detector corresponds to detector $D1$. The probability 
$p_{e,\bar{k},k}^m(n)$ can then be expressed as 
$p_{e,\bar{k},k}^m(n)=[p_{\bar{k},k,D1}^m(n)+p_{\bar{k},k,D_{Dc}}^m(n)/2]
(1-P_d)^{\bar{k}-n+1}+p_{\bar{k},k,D_{vac}}^m(n)(1-P_d)^{\bar{k}-n}P_d/2
+[p_{\bar{k},k,D1}^m(n)/2+p_{\bar{k},k,D_{Dc}}^m(n)/2](1-P_d)^{\bar{k}-n}
p_d(1-p_d)+[p_{\bar{k},k,D0}^m(n)/2+p_{\bar{k},k,D_{Dc}}^m(n)/2+
p_{\bar{k},k,D1}^m(n)](1-P_d)^{\bar{k}-n}
p_d(1-p_d)+[p_{\bar{k},k,D0}^m(n)+p_{\bar{k},k,D1}^m(n)+p_{\bar{k},k,D_{Dc}}^m(n)]
(1-P_d)^{\bar{k}-n}p_d^2/2$, where the parameters 
$p_{\bar{k},k,D0}^m(n)$, $p_{\bar{k},k,D1}^m(n)$ and $p_{\bar{k},k,D_{Dc}}^m(n)$
are given by Eq.~(\ref{pc0}), and $p_{\bar{k},k,D_{vac}}^m(n)$
is given by Eq.~(\ref{pcvac}). This quantity can be further simplified as
\begin{eqnarray}\label{ynsqh}
p_{e,\bar{k},k}^m(n)&=&(1-P_d)^{\bar{k}-n}\bigg[\bigg(1-\frac{p_d}{2}\bigg)p_{\bar{k},k,D1}^m(n) \nonumber\\
\ &+&\frac{p_dp_{\bar{k},k,D0}^m(n)}{2}+\frac{p_{\bar{k},k,D_{Dc}}^m(n)}{2} \nonumber\\
\ &+&p_d\bigg(1-\frac{p_d}{2}\bigg)p_{\bar{k},k,D_{vac}}^m(n)\bigg].
\end{eqnarray}

Similarly, the parameter $e_{vk}^m$, {\it i.e.}, the probability that Bob 
obtains an error when he receives a vacuum state and the
preceding signal is the state $\ket{\psi_k^m}$ 
has the form $e_{vk}^m=p_{e,0,k}^m(0)$, with $p_{e,\bar{k},k}^m(n)$ given by 
Eq.~(\ref{ynsqh}).

\section{Probabilities $dc_{vv}$, $dc_{vk}^m$ and 
$p_{dc,\bar{k},k}^m(n)$}\label{ap_e}

In this appendix we obtain an expression for the double click
probabilities $dc_{vv}$, $dc_{vk}^m$ and 
$p_{dc,\bar{k},k}^m(n)$ introduced in Sec.~\ref{dc_a_tds}.

The probability $dc_{vv}$, {\it i.e.}, the 
probability that Bob obtains a double click
when he receives from Eve a vacuum state and the preceding signal
is also a vacuum state is given by
\begin{equation}
dc_{vv}=p_d^2,
\end{equation}
where $p_d$ denotes again the dark count probability of Bob's
detectors.

The parameter $p_{dc,\bar{k},k}^m(n)$, {\it i.e.}, 
the probability that Bob obtains a double click in temporal mode $n$ and he 
does not obtain a click in any previous temporal mode $l$, with 
$n<l\leq{}\bar{k}$, when he receives  
the last $\bar{k}$ temporal modes of the signal $\ket{\psi_{k}^m}$
followed by one vacuum state is given by 
$p_{dc,\bar{k},k}^m(n)=p_{\bar{k},k,D_{Dc}}^m(n)
(1-P_d)^{\bar{k}-n+1}+p_{\bar{k},k,D_{vac}}^m(n)(1-P_d)^{\bar{k}-n}p_d^2
+[p_{\bar{k},k,D1}^m(n)+p_{\bar{k},k,D_{Dc}}^m(n)](1-P_d)^{\bar{k}-n}
p_d(1-p_d)+[p_{\bar{k},k,D0}^m(n)+p_{\bar{k},k,D_{Dc}}^m(n)](1-P_d)^{\bar{k}-n}
p_d(1-p_d)+[p_{\bar{k},k,D0}^m(n)+p_{\bar{k},k,D1}^m(n)+p_{\bar{k},k,D_{Dc}}^m(n)]
(1-P_d)^{\bar{k}-n}p_d^2$. This quantity can be further simplified as
\begin{eqnarray}\label{finquizas}
p_{dc,\bar{k},k}^m(n)&=&(1-P_d)^{\bar{k}-n}\bigg\{p_d\big[p_{\bar{k},k,D1}^m(n)+
p_{\bar{k},k,D0}^m(n)\big] \nonumber\\
\ &+&p_{\bar{k},k,D_{Dc}}^m(n)+p_d^2p_{\bar{k},k,D_{vac}}^m(n)\bigg\}.
\end{eqnarray}

Finally, the parameter $dc_{vk}^m$, {\it i.e.}, the probability that Bob 
obtains a double click when he receives a vacuum state and the
preceding signal is the state $\ket{\psi_k^m}$
has the form $dc_{vk}^m=p_{dc,0,k}^m(0)$, with $p_{dc,\bar{k},k}^m(n)$ given by 
Eq.~(\ref{finquizas}).


\bibliographystyle{apsrev}
\bibliographystyle{apsrev}

\begin{thebibliography}{99}
\bibitem{note1} In order to simplify our notation, from now on we
will employ the term ``signal state'' only to denote those light
pulses with a mean photon number bigger than zero. A light pulse
with an average photon number equal to zero will be always denoted
as a ``vacuum state''.
\bibitem{nsqp} In this section we shall use the parameter
$\tilde{d}$ defined as 
$\tilde{d}=\lceil{}t_df_c\rceil$ instead of the 
parameter $d$ employed in Sec.~\ref{susda}. The main motivation for this 
change of the notation is to try to avoid ambiguities when 
referring to the parameter $d$. In particular, since in the trusted device 
scenario we impose $d=0$ for the blocks of signal states 
illustrated in Fig.~\ref{strategy1_meas}.
\bibitem{dcr_uds} In this extended version of the DPS QKD protocol 
Alice and Bob could as well employ the double click rate information that 
is
measured in the experiment to try to discard those sequential attacks that 
highly decrease the value of the double click rate expected by the legitimate users. 
For instance, we have shown 
in Sec.~\ref{susda} that the eavesdropping strategy employed by Eve in the untrusted 
device scenario produces a double click rate equal to zero. 
However, if we consider a real implementation of the protocol,
then this last scenario is not so simple, since the double click rate 
expected by Alice and Bob due to the dark counts of the detectors is usually quite low.
\bibitem{dete_effi} In this section we consider that the parameter
$\eta_{det}$ also includes the losses due to Bob's interferometer.
\bibitem{nuevotexto} Note that in this definition we treat double clicks 
like single click events.
\end{thebibliography}

\end{document}